\documentclass[10pt,letterpaper]{article}
\usepackage[top=0.85in,left=2.75in,footskip=0.75in]{geometry}

\usepackage{amsmath,amssymb}

\usepackage{changepage}

\usepackage{textcomp,marvosym}

\usepackage{cite}

\usepackage{nameref,hyperref}

\usepackage[right]{lineno}

\usepackage[nopatch=eqnum]{microtype}
\DisableLigatures[f]{encoding = *, family = * }

\usepackage{array}

\newcolumntype{+}{!{\vrule width 2pt}}

\newlength\savedwidth

\raggedright
\usepackage[aboveskip=1pt,labelfont=bf,labelsep=period,justification=raggedright,singlelinecheck=off]{caption}

\makeatletter
\renewcommand{\@biblabel}[1]{\quad#1.}
\makeatother

\usepackage{lastpage,fancyhdr,graphicx}
\usepackage{epstopdf}
\usepackage{lmodern}
\fancyheadoffset[L]{2.25in}
\fancyfootoffset[L]{2.25in}
\usepackage[utf8]{inputenc}        
\usepackage[T1]{fontenc}           
\usepackage[dvipsnames, table]{xcolor}
\usepackage{tabularx}
\usepackage{multirow}
\usepackage{pifont}
\usepackage{csvsimple}
\usepackage[font={small},textfont={it},labelfont={bf}]{caption}
\usepackage{subcaption}
\usepackage{graphicx}
\usepackage{url}                   
\usepackage{booktabs}              
\usepackage{makecell}
\usepackage{amsfonts}              
\usepackage{amsmath}
\usepackage{nicefrac}              
\usepackage{microtype}             
\usepackage{enumitem}
\usepackage[export]{adjustbox}

\DeclareMathOperator*{\argmin}{arg\,min}
\newcommand{\indep}{\perp \!\!\! \perp}
\newtheorem{assumption}{Assumption}

\definecolor{dark_blue}{rgb}{0,0,.65}
\definecolor{dark_green}{rgb}{0,.5,.15}

\hypersetup{pdftex,  
  breaklinks=true,  
  colorlinks=true,
  linkcolor=dark_blue,
  citecolor=dark_green,
}
\colorlet{P}{ForestGreen}
\colorlet{I}{MidnightBlue}
\colorlet{C}{YellowOrange}
\colorlet{O}{DarkOrchid}
\colorlet{T}{Gray}

\begin{document}
\vspace*{0.2in}

\begin{flushleft}
  {\Large
    \textbf\newline{Step-by-step causal analysis of EHRs to ground decision-making} 
  }
  \newline
  \\
  Matthieu Doutreligne\textsuperscript{1,2,*},
  Tristan Struja\textsuperscript{3,4},
  Judith Abecassis\textsuperscript{1},
  Claire Morgand\textsuperscript{5},
  Leo Anthony Celi\textsuperscript{3,6,7},
  Gaël Varoquaux\textsuperscript{1}
  \bigskip
  \textbf{1} Inria, Soda, Saclay, France
  \\
  \textbf{2} Mission Data, Haute Autorité de Santé, Saint-Denis, France
  \\
  \textbf{3} Laboratory for Computational Physiology, Massachusetts Institute of Technology, Cambridge, MA 02139
  \\
  \textbf{4} Medical University Clinic, Division of Endocrinology, Diabetes \& Metabolism, Kantonsspital Aarau, Aarau, Switzerland
  \\
  \textbf{5} Agence Régionale de Santé Ile-de-France, France
  \\
  \textbf{6} Division of Pulmonary, Critical Care and Sleep Medicine, Beth Israel Deaconess Medical Center, Boston, MA 02215
  \\
  \textbf{7} Department of Biostatistics, Harvard T.H. Chan School of Public Health, Boston, MA 02115
  \bigskip

  %
  %





  * Corresponding author: m.doutreligne@has-sante.fr

\end{flushleft}
\section*{Abstract}

Causal inference enables machine learning methods to estimate treatment effects
of medical interventions from electronic health records (EHRs). The prevalence
of such observational data and the difficulty for randomized controlled trials (RCT) to cover all
population/treatment relationships make these methods increasingly attractive
for studying causal effects. However, researchers should be wary of many
pitfalls.

We propose and illustrate a framework for causal inference estimating the
effect of albumin on mortality in sepsis using an Intensive Care database
(MIMIC-IV) and comparing various sensitivity analyses to results from RCTs as gold-standard.

The first step is study design, using the target trial concept and the PICOT
framework: Population (patients with sepsis), Intervention (combination of
crystalloids and albumin for fluid resuscitation), Control (crystalloids only),
Outcome (28-day mortality), Time (intervention start within 24h of admission).
We show that too large treatment-initiation times induce immortal time bias.
The second step is selection of the confounding variables based on expert
knowledge. Increasingly adding confounders enables to recover the RCT results
from observational data. As the third step, we assess the influence
of multiple models with varying assumptions, showing that a doubly robust estimator (AIPW)
with random forests proved to be the most reliable estimator. Results show that
these steps are all important for valid causal estimates. A valid causal model
can then be used to individualize decision making: subgroup analyses showed that
treatment efficacy of albumin was better for patients >60 years old, males, and
patients with septic shock.

Without causal thinking, machine learning is not enough for optimal clinical
decision on an individual patient level. Our step-by-step analytic framework helps avoiding many pitfalls of applying machine learning to EHR data,
building models that avoid shortcuts and extract the best decision-making evidence.

\section*{Author summary}

Rich routine-care data, as EHR or claims, is useful to individualize decision
making using machine learning; but guiding interventions requires
causal inference. Unlike with an RCT, interventions in routine data do
not easily enable an apple-to-apple measure of the effect of an
intervention, leading to many analytical pitfalls, particularly in
time-varying data. We study these in a tutorial spirit, making
the code and data openly available. We give 5 analytical steps for
data-driven individualized interventions: Step
1) Study design, where common pitfalls are selection bias, with information
unequally collected across treatment and control patients, and immortal time
bias, where the inclusion-defining event interacts with the
intervention time. Step 2) Identification of the causal assumptions and
categorization of confounders. Step 3) Estimation of the causal effect of
interest by correct aggregation of confounders and selection of an appropriate
statistical model. Step 4) Assessing the
analysis' robustness to assumptions, and finally Step 5) Individualizing
treatment decision, by exploring treatment heterogeneity, eg across subgroups.
Studying choice of fluid resuscitation in sepsis, we show that common
mistakes in steps 1, 2, and 3 equally compromise causal validity.

\linenumbers

\section*{Introduction: data-driven decisions require causal inference}

Informing a care option extends beyond merely predicting the occurrence
of an event; it involves estimating the effect of the corresponding
treatment effects. Routine-care data comes naturally to mind to guide
routine decisions, but they require care to estimate treatment effects as
they are observational, unlike Randomized controlled trials (RCTs). This
context calls for causal inference statistical frameworks.
But merely applying these tools to the data does suffice to ensure the
validity of the inferences; numerous considerations must be carefully addressed.

\paragraph{Individualized Medicine and Machine Learning Challenges}
Machine learning plays a pivotal role in individualized medicine  \cite{rajkomar2018scalable,liu2019comparison,li2020behrt,beaulieu2021machine,aggarwal2021diagnostic}. It
demonstrated superior performance over traditional rule-based clinical scores in predicting a patient's readmission risk, mortality, or future comorbidities using Electronic Health Records (EHRs) \cite{rajkomar2018scalable,liu2019comparison,li2020behrt,beaulieu2021machine,aggarwal2021diagnostic}.
However, mounting evidence suggests that machine-learning models can inadvertently perpetuate and exacerbate biases present in the data \cite{rajkomar2018ensuring}, including gender or racial biases \cite{singh2022generalizability,gichoya2022ai}, and the marginalization of under-served populations \cite{seyyed2021underdiagnosis}. These biases are typically encoded by capturing shortcuts—stereotypical or distorted features in the data \cite{geirhos2020shortcut,winkler2019association,degrave2021ai}.
For instance, numerous machine learning algorithms rely on post-treatment information \cite{badgeley2019deep,obermeyer2019dissecting,yuan2021temporal,wong2021external}, exemplified by a diagnostic model for skin cancer that depends on surgical marks \cite{winkler2019association}. For Intensive Care Unit data, focus of our study, such information markedly improves mortality prediction (Figure S1 \ref{apd:fig:motivating_example}), but cannot inform decisions.

\paragraph{The Importance of Causal Reasoning in Data-Driven Decision-Making} \cite{prosperi2020causal}
While conventional machine learning relies on retrospective to generate predictions of future effects \cite{plecko2022causal}, truly informing decision-making needs a comparison of potential outcomes with and without the intervention. This involves estimating a causal effect, mirroring the methodology employed in RCTs \cite{prosperi2020causal}. However, RCTs encounter challenges such as selection biases \cite{travers2007external,averitt2020translating}, difficulties in recruiting diverse populations, and limited sample sizes for exploring treatment heterogeneity across subgroups. Routinely collected data presents a unique opportunity to assess real-life benefit-risk trade-offs associated with a decision \cite{desai2021broadening}, with reduced sampling bias and sufficient data to capture heterogeneity \cite{rekkas2023standardized}. Nevertheless, estimating causal effects from such data is challenging due to the confounding of the intervention by indication. Therefore, dedicated statistical techniques are imperative to emulate a "target trial" \cite{hernan2016specifying} from observational data.

\paragraph{Multiple Perspectives on Evidence-based Decision Making}
Across different fields, existing literature has emphasized different challenges associated
with estimating treatment effects using observational data. While epidemiologic
studies underscore the importance of the target trial approach
\cite{von2007strengthening,benchimol2015reporting,hernan2020causal,schneeweiss2021conducting,zeng2022uncovering},
there emphasis primarily lies on biases that arise from temporal effects \cite{suissa2008immortal,Oke2021leadtimebias,fu2021timing,hernan2016specifying,wang2022understanding,Bankhead2017attritionbias} or confounding variables \cite{greenland1999causal,vanderweele2019principles,loh2021confounder}, with
relatively less attention to issues arising from estimator selection.
Recent replications of RCTs using observational data did not explore the impact
of modern machine learning methods on the robustness of the results
\cite{schneeweiss2021conducting,wang2023emulation}.

In contrast, machine learning and causal inference literature predominantly
studies estimators
\cite{belloni2014high,chernozhukov2018double,shalit2016tutorial,sharma2018tutorial,moraffah2021causal}
: propensity score matching \cite{stuart2010matching}, inverse probability
weighting \cite{austin2015moving}, outcome models \cite{robins1986role}, doubly
robust methods, \cite{chernozhukov2018double} or deep learning based models
\cite{johansson2022generalization}. This literature may be opaque for some due to intricate mathematical details and unverifiable assumptions. Guidelines seldom
address time-related biases, or covariate aggregation which
frequently emerge in datasets with temporal dependencies
\cite{suissa2008immortal,fu2021timing}. Recently, the machine learning community
shifted its focus from EHR data to simulated data, which may not capture the
complexities of real-world data \cite{schuler2017targeted,dorie2019automated,
  alaa2019validating, curth2021really}.

In this work, we bring together epidemiological concepts and
principles from statistical and machine learning literature. We adopt
an empirical perspective to answer practical needs of applied researchers.
A study of choices spread out across the analysis
--study design, consideration of
confounders, and selection of estimators (refer to Section
\nameref{sec:inference_flow})-- highlights their equal importance in ensuring
the validity of results. To illustrate and compare biases, we investigate
the impact of albumin on sepsis mortality using data from a publicly available
intensive care database, MIMIC-IV \cite{johnson2020mimic} (section
\nameref{sec:application_on_mimic_iv}).

The primary focus of the main section is on accessibility, with technical
details expanded in the appendices.

\section*{Step-by-step framework for robust decision-making from EHR data}\label{sec:inference_flow}

\begin{figure*}[t!]
  \centering
  \includegraphics[width=0.9\linewidth]{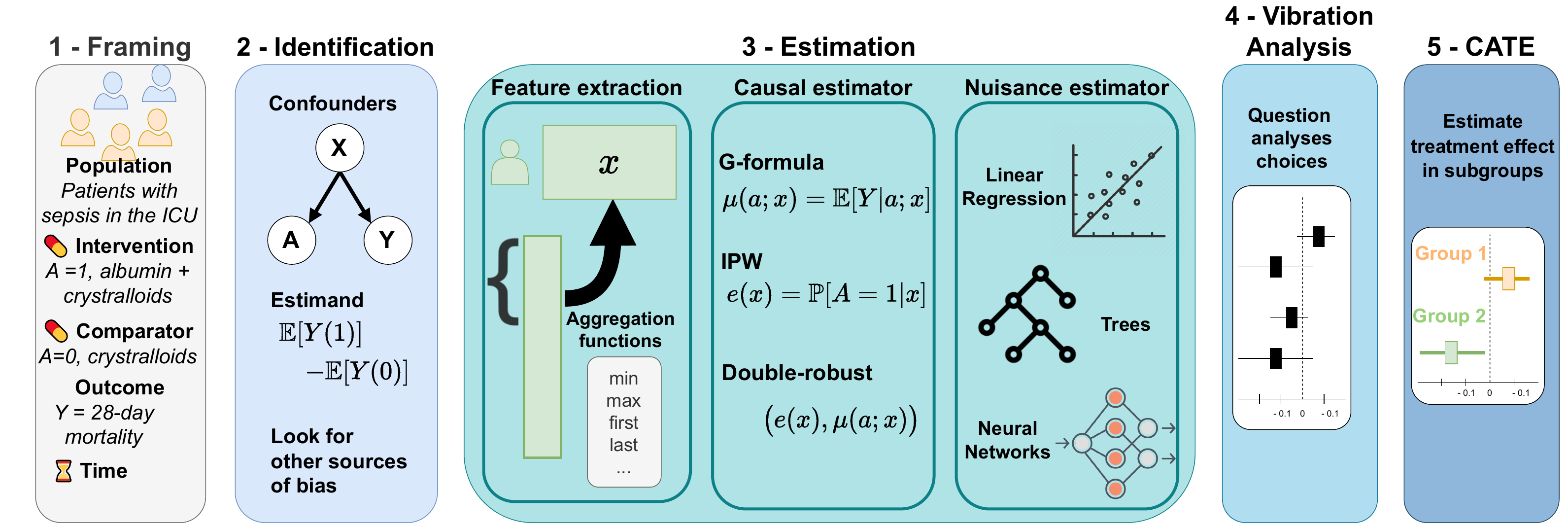}
  \caption{\textbf{Step-by-step analytic framework} -- The complete
    inference pipeline confronts the analyst with
    many choices, some guided by domain knowledge, others
    by data insights. Making those choices explicit is necessary to ensure
    robustness and reproducibility.}\label{fig:inference_framework}
\end{figure*}

Whether or not using machine learning, many pitfalls threaten an analysis'
value for decision-making. To avoid these pitfalls, we outline a simple
step-by-step analytic framework illustrated in Figure
\ref{fig:inference_framework} for retrospective case-control studies. We frame
the medical question as a target trial \cite{hernan2021methods} to match the
design to an RCT giving the gold standard average effect. Then we probe for
heterogeneity --predictions on sub-groups--  going beyond what RCTs can
achieve.

\subsection*{Step 1: study design -- Frame the question to avoid biases}\label{sec:framing}

\begin{table*}[b!]
  \resizebox{\linewidth}{!}{%
    \begin{tabular}{|l|l|l|l|}
      \hline
      \multicolumn{1}{|c|}{\textbf{PICO component}} & \multicolumn{1}{c|}{\textbf{Description}}       & \multicolumn{1}{c|}{\textbf{Notation}}                 & \multicolumn{1}{c|}{\textbf{Example}}         \\ \hline
      \cellcolor{P}\textbf{Population}              & What is the target population of interest?      & $X \sim \mathbb{P}(X)$, the covariate distribution     & \makecell[l]{Patients with sepsis in the ICU} \\ \hline
      \cellcolor{I}\textbf{Intervention}            & What is the treatment?                          & \makecell[l]{$A \sim \mathbb{P}(A=1)=p_A$,                                                             \\ the probability to be treated} & Combination of crystalloids and albumin \\ \hline
      \cellcolor{C}\textbf{Control}                 & What is the clinically relevant comparator?     & $1-A \sim 1-p_A$                                       & Crystalloids only                             \\ \hline
      \cellcolor{O}\textbf{Outcome}                 & \makecell[l]{What are the outcomes to compare?} & \makecell[l]{$Y(1), Y(0) \sim \mathbb{P}(Y(1), Y(0))$,                                                 \\ the potential outcomes distribution}  & 28-day mortality \\ \hline
      \cellcolor{T}\textbf{Time}                    & \makecell[l]{Is the start of follow-up aligned                                                                                                           \\with intervention assignment?} & \makecell[l]{N/A}  & \makecell[l]{Intervention \\ within the first day} \\ \hline
    \end{tabular}%
  }\\
  \caption{PICO(T) components help to clearly define the
    medical question of interest.}\label{table:picot}
\end{table*}

Grounding decisions on evidence needs well-framed questions, defined by their
PICO(T) components. Population, Intervention, Control, and Outcome
\cite{richardson1995well,riva2012your}, and in case of EHRs or claims data an
additional time component, are necessary to concord with a (hypothetical) target
randomized clinical trial \cite{hernan_using_2016,wang2023emulation} -- Table
\ref{table:picot}. A selection flowchart such as in
\nameref{apd:selection_flowchart} makes inclusion and exclusion choices for PICOT explicit.

Without care in defining these PICO(T) components, non-causal associations
between treatment and outcomes can easily be introduced into an analysis
\cite{catalogofbias}. The time-varying nature of EHR calls for checking
systematically of the Population and Time components by addressing two commonly
encountered types of bias.

\paragraph{Selection Bias:} In EHRs, outcomes and treatments are often not
directly available and need to be inferred from indirect events. These signals
could be missing not-at random, sometimes correlated with the treatment
allocation \cite{weiskopf2023healthcare}. For example, billing codes can be
strongly associated with case-severity and cost. Consider comparing the
effectiveness of fluid resuscitation with albumin to crystalloids. As albumin is
more costly, this treatment is more likely to have a sepsis billing code. On the
contrary, for patients treated with crystalloids, only the most severe cases
will have a billing code. Naively comparing patients would overestimate the
effect of albumin.

\paragraph{Immortal time bias:} Improper alignment of the inclusion defining
event and the intervention time is a major source of bias in time-varying data
\cite{suissa2008immortal,hernan2016specifying,wang2022understanding}. Immortal time bias (illustrated in
Appendix \ref{apd:fig:immortal_time_bias}) occurs when the follow-up period,
i.e. cohort entry, starts before the intervention, e.g. prescription for a
second-line treatment. In this case, the treated group will be biased towards
patients still alive at the time of assignment and thus overestimating the
effect size. Other frequent temporal biases are lead time bias
\cite{Oke2021leadtimebias,fu2021timing} or right censorship  \cite{hernan2016specifying}, and attrition bias
\cite{Bankhead2017attritionbias}. Good
practices include explicitly stating the cohort inclusion event   \cite[Chapter~10:Defining Cohorts]{ohdsi2019book} and defining
an appropriate grace period between starting time and the intervention
assignment \cite{hernan2016specifying}. At this step, a population timeline can
help.

\subsection*{Step 2: identification -- List necessary information to answer the causal question}\label{sec:identification}

The identification step builds a causal model to answer the research question.
Indeed, the analysis must compensate for differences between
treated and non-treated that are not due to the intervention
(\cite[chapter~1]{pearl2018book},
\cite[chapter~1]{hernan2020causal}).

\paragraph{Causal Assumptions}

Valid causal inference requires assumptions  \cite{rubin2005causal} --detailed
in \nameref{apd:causal_assumptions}. The analyst should thus review the
plausibility of the following: 1) Unconfoundedness: after adjusting for the
confounders as ascertained by domain expert insight, treatment allocation should
be random; 2) Overlap --also called positivity-- the distribution of confounding
variables overlaps between the treated and controls --this is the only
assumption testable from data \cite{austin2015moving}--; 3) No interference
between units and consistency in the treatment, a reasonable assumption in most
clinical questions.

\paragraph{Categorizing covariates}
Potential predictors --covariates-- should be categorized depending on their
causal relations with the intervention and the outcome (illustrated in
\nameref{apd:causal_variables}): \emph{confounders} are common causes of the
intervention and the outcome; \emph{colliders} are caused by both the
intervention and the outcome; \emph{instrumental variables} are a cause of the
intervention but not the outcome, \emph{mediators} are caused by the
intervention and is a cause of the outcome. Finally, \emph{effect modifiers}
interact with the treatment, and thus
modulate the treatment effect in subpopulations \cite{attia2022proposal}.

To capture a valid causal effect, the analysis should only include confounders
and possible treatment-effect modifiers to study the resulting heterogeneity.
Regressing the outcome on instrumental and post-treatment variables (colliders
and mediators) will lead to biased causal estimates
\cite{vanderweele2019principles}. Drawing causal Directed Acyclic Graphs
(DAGs) \cite{greenland1999causal}, \emph{eg} with a webtool such as DAGitty \cite{textor2011dagitty}, helps capturing the relevant
variables and defining a suitable estimand or effect measure.

Unconfoundedness --inclusion of all confounders in the analysis-- is a strong
assumption that can be difficult to ascertain in practice applications. In these cases,
sensitivity analyses for omitted variable bias allow to test the robustness of
the results to missing confounders \cite{cinelli2020making}, proximal inference
can be used to leverage proxy of unobserved confounders
\cite{tchetgen2024introduction}, and the presence of a natural experiment or RCT might
identify the desired causal effect without unconfoundedness \cite[Chapter 5,
  9]{wager2020stats}.

The
\emph{estimand} is the final causal quantity estimated from the data.
Depending on the question, different estimands are better suited to contrast
the two potential outcomes E[Y(1)] and E[Y(0)] \cite{imbens_nonparametric_2004,colnet2023risk}. For continuous outcomes, risk
difference is a natural estimand, while for binary outcomes (e.g. events) the
choice of estimand depends on the scale. Whereas the risk difference is very
informative at the population level, e.g. for medico-economic decision-making,
the risk ratio and the hazard ratio are more informative at the level of
sub-groups or individuals \cite{colnet2023risk}.

\paragraph{Causal estimators}

A given estimand can be estimated through different methods. One can model the
outcome with regression models also known as
G-formula, \cite{robins1986role} and use it as a predictive counterfactual model
for all possible treatments for a given patient. Alternatively, one can model
the propensity of being treated for use in matching or Inverse Propensity
Weighting (IPW) \cite{austin2015moving}. Finally, doubly robust methods model
both the outcome and the treatment, benefiting from the convergence of both
models \cite{wager2020stats}. There is a variety of doubly robust models,
reviewed in \nameref{apd:causal_estimators}.

\subsection*{Step 3: Statistical estimation -- Compute the causal effect of interest}\label{sec:estimation}

%

\paragraph{Confounder aggregation}
Confounders captured via measures collected over multiple time points must be
aggregated at the patient level. Simple forms of aggregation include taking the
first or last value before a time point, or an aggregate such as mean or median
over time. More elaborate choices may rely on hourly aggregations providing more
detailed information on the disease course such as vital signs. They may reduce
confounding bias between rapidly deteriorating and stable patients but also
increase the number of confounders making estimation more challenging
\cite{damour2020overlap}. The increase of variance occurs either in
arbitrarily small propensity scores for treatment models or in hazardous
extrapolation from one group to another for outcome model. If multiple
choices appear reasonable, one should compare them in a vibration analysis
(see \nameref{sec:vibration_analysis}).

Beyond tabular data, unstructured clinical text may capture confounding or
prognostic information \cite{horng2017creating,jiang2023health} which can be
added in the causal model \cite{zeng2022uncovering}.
However, high-dimensional
confounder space such as text may break the positivity assumption just as hourly
aggregation choices for measurements.

Missing covariate values might also be a source of confounding. Some statistical
estimators (such as forests) can directly incorporate them as supplementary
covariates. Others, such as linear models, require imputations.
\nameref{apd:statistical_considerations} details general sanity checks for
imputation strategies when using statistical estimators.

\paragraph{Statistical estimation models of outcome and treatment}

The causal estimators use models of the outcome or the treatment --called
nuisances. There is currently no clear best practice to choose the corresponding
statistical model \cite{wendling2018comparing, dorie2019automated}. The
trade-off lies between simple models risking misspecification of the nuisance
parameters versus flexible models risking to overfit the data at small sample
sizes. Stacking models of different complexity as in a super-learner is a good
solution to navigate the trade-off \cite{van2007super,doutreligne2023select}.

\subsection*{Step 4: Vibration analysis -- Assess the robustness of the hypotheses}\label{sec:vibration_analysis}

Some choices in the pipeline may not be clear cut. Several options should then
be explored, to derive conceptual error bars going beyond a single statistical
model. When quantifying the bias from unobserved confounders, this process is
sometimes called sensitivity analysis \cite{schneeweiss2006sensitivity,thabane2013tutorial,fda_statistical_2021}.
Following \cite{patel2015assessment}, we use the term vibration analysis to
describe the sensitivity of the results to all analytic choices.

\subsection*{Step 5: Treatment heterogeneity -- Compute treatment effects on subpopulations}\label{sec:treatment_heterogeneity}

Once the causal design and corresponding estimators are established, they can be
used to explore the variation of treatment effects among subgroups. A
causally-grounded model can be used to predict the effect of the treatment from
all the covariates --confounders and effect modifiers-- the \emph{Conditional Average
  Treatment Effect} (CATE) \cite{robertson2021assessing}. Practically, CATEs can be estimated by regressing
an individual's predictions given by the causal estimator against the sources of
heterogeneity (details in \nameref{apd:cate_results}).

\section*{Application: evidence from MIMIC-IV on which resuscitation fluid to use}%
\label{sec:application_on_mimic_iv}

We now use the above framework to extract evidence-based decision rules for
resuscitation. Ensuring optimal organ perfusion in patients with septic shock
requires resuscitation by reestablishing circulatory volume with intravenous
fluids. While crystalloids are readily available, inexpensive and safe, a
large fraction of the administered volume is not retained in the vasculature.
Colloids offer the theoretical benefit of retaining more volume, but might be
more costly and have adverse effects \cite{annane2013effects}. Meta-analyses
from multiple pivotal RCTs found no effect of adding albumin to crystalloids
\cite{xu2014comparison,li2020resuscitation} on 28-day and 90-day mortality.  Given this
previous evidence, we thus expect no average effect of albumin on mortality in
sepsis patients. However, studies --RCT \cite{caironi2014albumin}  and observational
\cite{zhou2021early}-- have
found that septic-shock patients do benefit from albumin.

\paragraph{Emulated trial: Effect of albumin in combination with crystalloids
  compared to crystalloids alone on 28-day mortality in patients with sepsis}\label{emulated_trial}

\begin{figure}[h!]
  \centering
  \includegraphics[width=\linewidth]{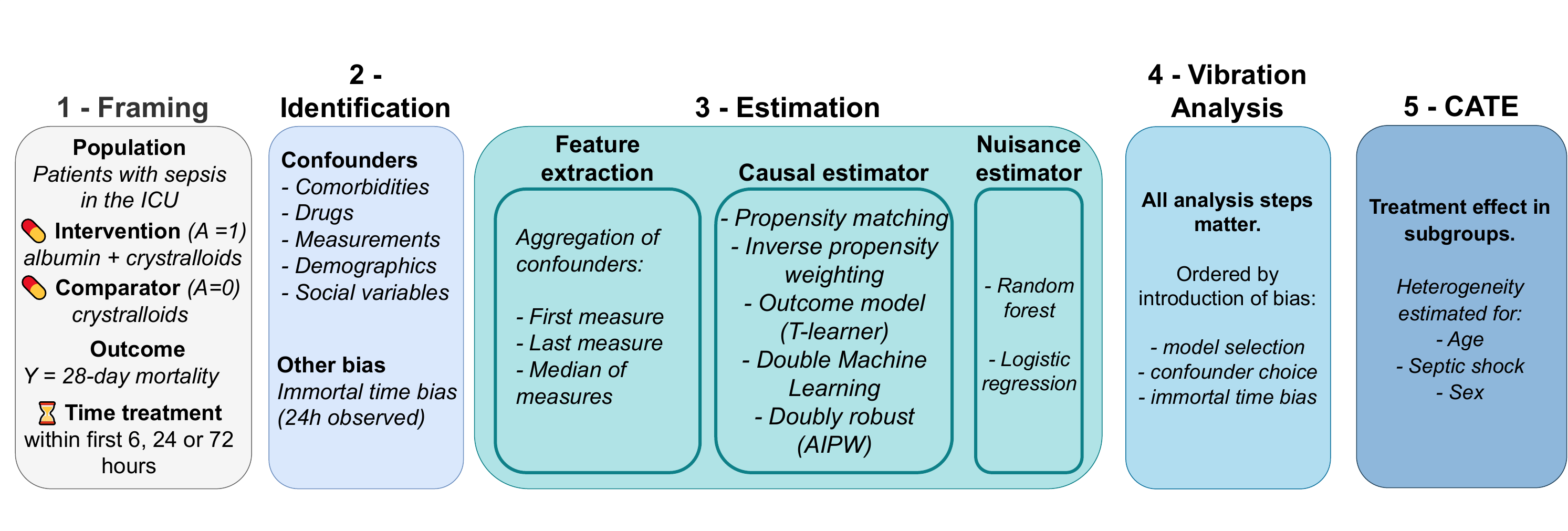}
  \caption{\textbf{Application of the step-by-step framework on which resuscitation fluid to use.}}\label{fig:study_summary}
\end{figure}

Multiple published RCTs can validate the analysis pipeline before
investigating sub-population effects for individualized decisions. Using
MIMIC-IV \cite{johnson2020mimic}, we compare the magnitude of biases introduced by reasonable
choices in the different analytical steps recalled in Figure \ref{fig:study_summary}.

MIMIC-IV is a publicly available database that contains information from real
ICU stays of patients admitted to one tertiary academic medical center, Beth
Israel Deaconess Medical Center (BIDMC), in Boston, United States between 2008
and 2019. The data in MIMIC-IV has been previously de-identified, and the
institutional review boards of the Massachusetts Institute of Technology (No.
0403000206) and BIDMC (2001-P-001699/14) both approved the use of the database
for research. The database contains comprehensive information from ICU stays
including vital signs, laboratory measurements, medications, and mortality data
up to one year after discharge.

\subsection*{Step 1: Study design -- effect of crystalloids on mortality in sepsis}%
\label{sec:framing_mimic_iv}

\begin{itemize}[leftmargin=2ex]
  \item \textcolor{P}{Population}: Patients with sepsis in an ICU stay according
        to the sepsis-3 definition. Other inclusion criteria: sufficient
        follow-up of at least 24 hours, and age over 18 years.
        \nameref{apd:selection_flowchart} details the selection flowchart and
        \nameref{apd:albumin_for_sepsis:table1_complete} the population
        characteristics.

  \item \textcolor{I}{Intervention}: Treatment with a combination of
        crystalloids and albumin during the first 24 hours of an ICU stay.

  \item \textcolor{C}{Control}: Treatment with crystalloids only in the first 24
        hours of an ICU stay.

  \item \textcolor{O}{Outcome}: 28-day mortality.

  \item \textcolor{T}{Time}: Follow-up begins after the first administration of
        crystalloids. Thus, we potentially introduce a small immortal time bias
        by allowing a time gap between follow-up and the start of the albumin
        treatment --see the full timeline in \nameref{apd:graphical_timeline}. Because
        we are only considering the first 24 hours of an ICU stay, we
        hypothesize that this gap is insufficient to affect our results. We test
        this hypothesis in the vibration analysis step.
\end{itemize}

In MIMIC-IV, these inclusion criteria yield 18,121 patients of which 3,559 were treated with a combination of crystalloids and albumin. While glycopeptide antibiotic therapy   was similar between both groups (51.8\% crystalloid vs 51.5\% crystalloids + albumin), aminoglycosides, carbapenems, and beta-lactams were more frequent in the crystalloid only group (2.0\% vs. 0.7\%, 4.3\% vs. 2.6\%, and 35.5\% vs. 13.8\%, respectively). The crystalloid only group was more frequently admitted as an emergency (57.3\% vs. 30.7\%). Vasopressors (80.2\% vs 41.7\%) and ventilation (96.8\% vs 87.0\%) were more prevalent in the treated populations, underlying the overall higher severity of patients receiving albumin (mean SOFA at admission 6.9 vs. 5.7).
Table \ref{table:albumin_for_sepsis:table1_simple} details
patient characteristics.

\begin{table}[h!]
  \centering\small
  \resizebox{\columnwidth}{!}{
    \begin{tabular}{llllll}
\toprule
{} & Missing &       Overall & Cristalloids only & Cristalloids + Albumin & P-Value \\
                           &         &               &                   &                        &         \\
\midrule
n                          &         &         18421 &             14862 &                   3559 &         \\
Female, n (\%)              &         &   7653 (41.5) &       6322 (42.5) &            1331 (37.4) &         \\
White, n (\%)               &         &  12366 (67.1) &       9808 (66.0) &            2558 (71.9) &         \\
Emergency admission, n (\%) &         &   9605 (52.1) &       8512 (57.3) &            1093 (30.7) &         \\
admission\_age, mean (SD)   &       0 &   66.3 (16.2) &       66.1 (16.8) &            67.3 (13.1) &  <0.001 \\
SOFA, mean (SD)            &       0 &     6.0 (3.5) &         5.7 (3.4) &              6.9 (3.6) &  <0.001 \\
lactate, mean (SD)         &    4616 &     3.0 (2.5) &         2.8 (2.4) &              3.7 (2.6) &  <0.001 \\
\bottomrule
\end{tabular}
  }
  \\[.5ex]

  \caption{Characteristics of the trial population measured on the first 24
    hours of ICU stay. Appendix \ref{apd:table:albumin_for_sepsis:table1_complete}
    describes all confounders used in the analysis.}\label{table:albumin_for_sepsis:table1_simple}
\end{table}

\subsection*{Step 2: Identification -- listing confounders}\label{sec:identification_mimic_iv}

For confounders selection we use a causal DAG shown in Figure
\nameref{apd:causal_diagram_albumin}. Gray confounders are not controlled for since
they are not available in the data. However, resulting confounding biases are
captured by proxies such as comorbidity scores (SOFA or SAPS II) or other
variables (eg. race, gender, age, weight).
\nameref{apd:albumin_for_sepsis:table1_complete} details confounders
summary statistics for treated and controls.

\paragraph{Causal estimators:}

We implemented multiple estimation strategies, including Inverse Propensity
Weighting (IPW), outcome modeling (G-formula) with T-Learner, Augmented Inverse
Propensity Weighting (AIPW) and Double Machine Learning (DML). We used the
python packages dowhy \cite{sharma2018tutorial} for IPW implementation and
EconML \cite{battocchi2019econml} for all other estimation strategies.
Confidence intervals were estimated by bootstrap (50 repetitions).
\nameref{apd:causal_estimators} and
\nameref{apd:packages} detail the estimators and the available Python
implementations. \nameref{apd:statistical_considerations} details statistical considerations that we identified as important but missing in these packages, namely lack of cross fitting estimators, bad practices for imputation, or lack of closed form confidence intervals.

\subsection*{Step 3: Statistical estimation}\label{sec:estimation_mimic_iv}

\paragraph{Confounder aggregation:}

We tested multiple aggregations such as the last value before the start of the
follow-up period, the first observed value, and both the first and last values
as separated features. Missing values were median imputed for numerical
features, categorical variables were one-hot encoded (thus discarding missing values).

\paragraph{Outcome and treatment estimators:}

To model the outcome and treatment, we used two common but different
estimators: random forests and ridge logistic regression implemented with scikit-learn
\cite{pedregosa2011scikit}. We chose the
hyperparameters with a random search procedure (\nameref{apd:hyper_parameter_search}). While logistic regression handles
predictors in a linear fashion, random forests bring the benefit of modeling
non-linear relations.

\subsection*{Step 4: Vibration analysis -- Comparing sources of systematic errors}%
\label{sec:vibration_analysis_mimic_iv}

\paragraph{Study design flaw -- Illustration of immortal time bias:}

\begin{figure*}[h!]
  \begin{subfigure}[b]{\linewidth}
    \caption{\bfseries Framing -- Immortal Time Bias}\label{fig:vibration:itb}
    \includegraphics[width=0.765\linewidth, right]{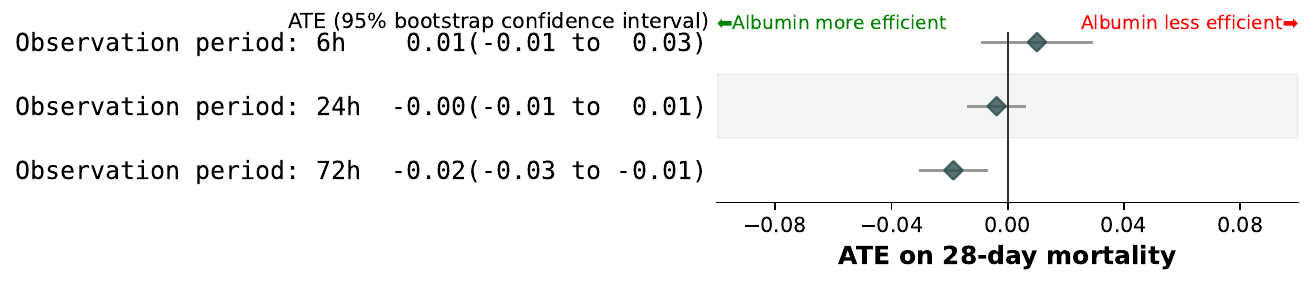}
  \end{subfigure}
  \vfill
  \begin{subfigure}[b]{\linewidth}
    \centering
    \caption{\bfseries Identification -- confounders choice}\label{fig:vibration:confounders}
    \includegraphics[width=.9\linewidth, right]{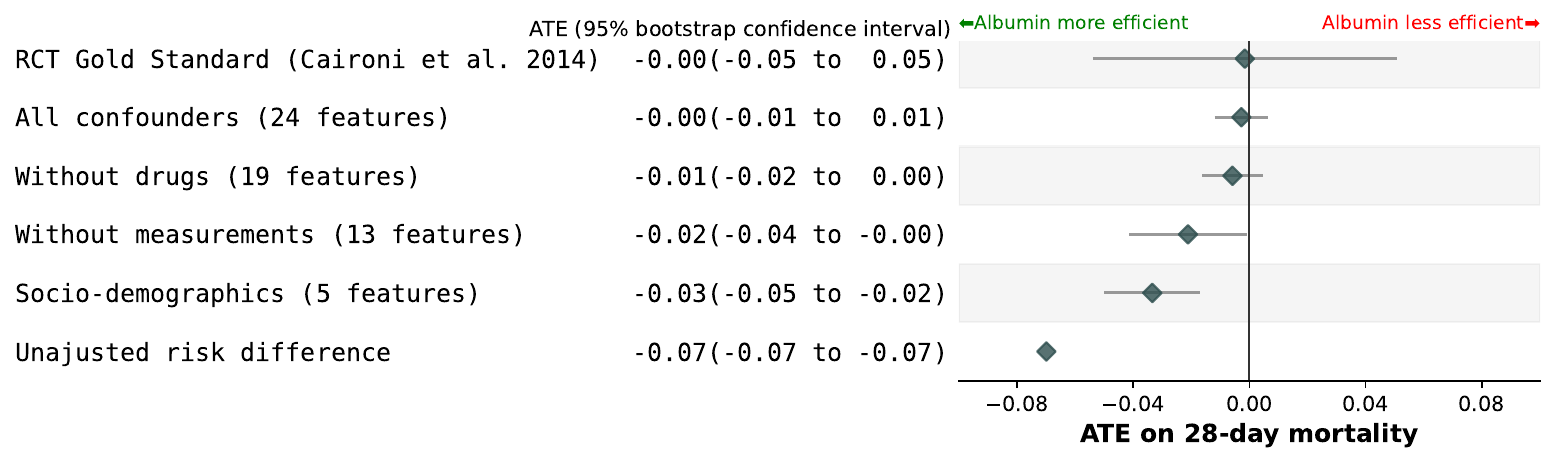}
  \end{subfigure}
  \vfill
  \begin{subfigure}[b]{\linewidth}
    \centering
    \caption{\bfseries Model selection}\label{fig:vibration:models}
    \includegraphics[width=0.891\linewidth, right]{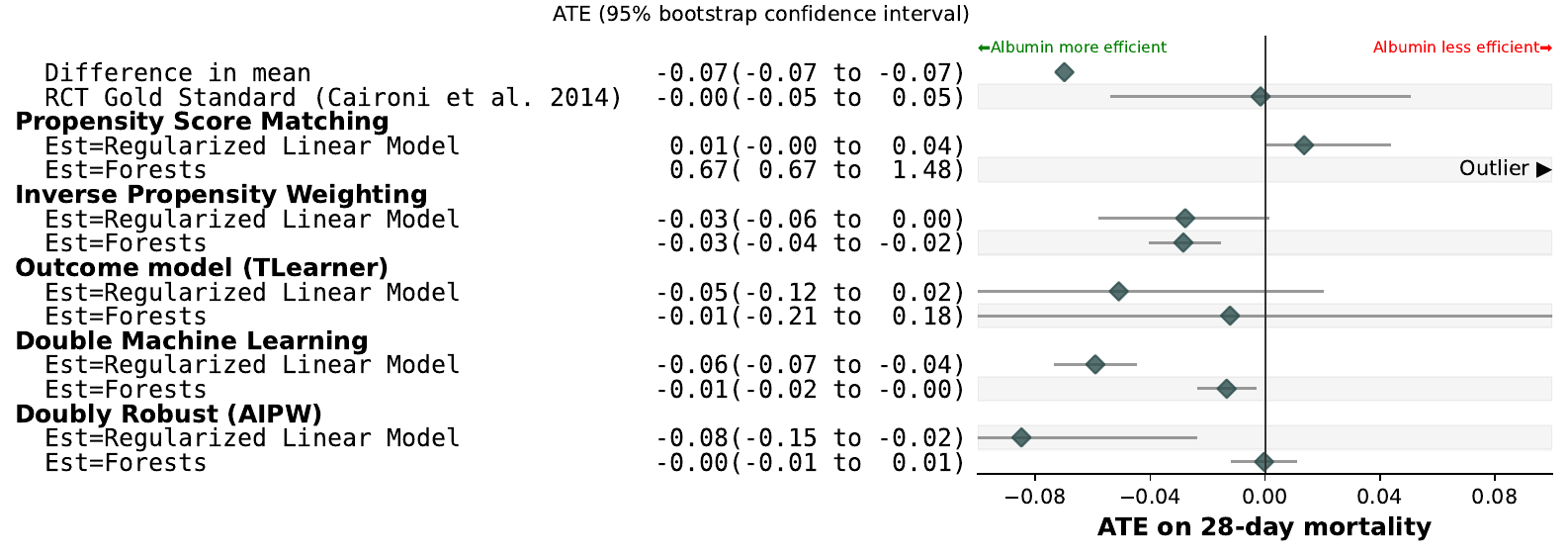}
  \end{subfigure}
  \vfill
  \caption{\textbf{The effect of choices on the three
      analytical steps} -- All three
    analytical steps are equally important for the validity of the analysis.
    \ref{fig:vibration:itb}) Framing step: Poor framing introduces time bias: A longer
    observation period (72h) artificially favors the efficacy of Albumin.
    \ref{fig:vibration:confounders}) Identification step: Choosing less informed confounders set
    introduces increasing bias in the results. \ref{fig:vibration:models}) Model selection step:
    Different estimators give different results. Score matching yields
    unconvincingly high estimates, inconsistent with the published RCT. With
    other causal approaches, using linear estimators for nuisances suggest a
    reduced mortality risk for albumin, while using forests for nuisance models
    points to no effect, which is consistent with the RCT gold standard.
    \\The diamonds depict the mean effect and the bar are the 95\% confidence
    intervals obtained respectively by 30, 30 and 50 bootstrap repetitions. For
    framing and identification, the estimator is a doubly robust learner (AIPW)
    with random forests for nuisances. Features are aggregated by taking the
    first and last measurements for all experiments.}\label{fig:vibration_analysis}
\end{figure*}

To illustrate the risk of immortal-time bias, we vary the eligibility period of
treatment or control in a shorter or longer time window than 24 hours. As
explained in \autoref{sec:framing}, a longer eligibility period means that
patients are more likely to be treated if they survived up to the intervention
and hence the study is biased to overestimate the beneficial effect of the
intervention. Figure \ref{fig:vibration:itb} shows that longer eligibility
periods lead to albumin being markedly more efficient (detailed results with causal forest and other choices of aggregation in \nameref{apd:detailed_results_itb}).

\paragraph{Confounder choice flaw} We consider other choice of confounding variables
(\nameref{apd:vibration_analysis_for_confounders}). Figure
\ref{fig:vibration:confounders} shows that a less thorough choice, neglecting
the administrated drugs, makes little to no difference. Major errors, such as
omitting the biological measurements or using only socio-demographical
variables, lead to sizeable bias. This is consistent with the literature
highlighting the importance of a clinically valid DAG \cite{greenland1999causal}.

\paragraph{Estimation choices flaw -- Confounder aggregation, causal and nuisance estimators:}

Figure \ref{fig:vibration:models} shows varying confidence intervals (CI)
depending on the method. Doubly-robust methods provide the narrowest CIs,
whereas the outcome-regression methods have the largest CI. The estimates of the
forest models are closer to the consensus across prior studies (no effect) than
the logistic regression indicating a better fit of non-linear relationships. We
only report the first and last pre-treatment feature aggregation strategy, since
detailed analysis showed little differences for other aggregations (\nameref{apd:detailed_results} for complete results, and \nameref{apd:vibration_analysis_for_aggregation} for a detailed study on aggregation choices). Both methodological studies \cite{naimi2023challenges} and
consistency with published RCTs suggest to prefer doubly-robust approaches.

\subsection*{Step 5: Treatment heterogeneity -- Which treatment for a sub-population?}%
\label{sec:treamtent_heterogeneity_mimic_iv}

\begin{figure}[h!]
  \begin{minipage}{.4\linewidth}
    \caption{\textbf{Subgroup distributions of Individual Treatment
        effects}:
      better treatment efficacy for patients older than 60 years, septic shock,
      and to a lower extent males. The final estimator is ridge regression. The
      boxes contain the $25^\text{th}$ and $75^\text{th}$ percentiles of the CATE
      distributions with the median indicated by the vertical line. The whiskers
      extend to 1.5 times the inter-quartile range of the
      distribution.}\label{fig:albumin_for_sepsis:cate_results}
  \end{minipage}%
  \hfill%
  \begin{minipage}{.6\linewidth}
    \includegraphics[width=\linewidth]{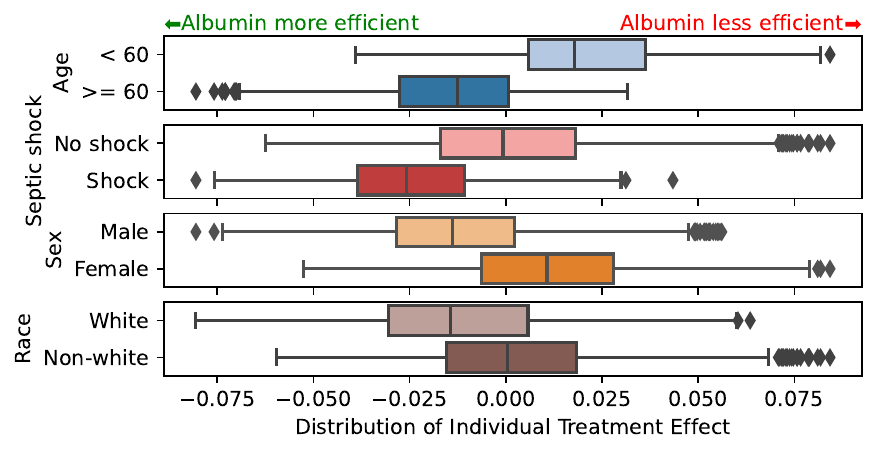}
  \end{minipage}%
\end{figure}

With adequate choice of study design, confounding variables and causal
estimator, the average treatment effect matches well published findings:
Pooling evidence from high-quality RCTs, no effect of albumin in severe
sepsis was demonstrated for both 28-day mortality (odds ratio (OR) 0.93,
95\% CI 0.80-1.08) and 90-day mortality (OR 0.88, 95\% CI 0.761.01)
\cite{xu2014comparison}.
Having validated the analytical pipeline, we can use it to inform
decision-making. We explore
heterogeneity along four binary patient characteristics, displayed in Figure
\ref{fig:albumin_for_sepsis:cate_results}. We find that albumin is beneficial
with patient with septic shock consistent with one
RCT  \cite{caironi2014albumin}. It is also beneficial for older patients (age >=60) and males. \nameref{apd:hte} details the heterogeneity analysis.

\section*{Discussion and conclusion}\label{sec:discussion}


Valid decision-making evidence from EHR data requires a clear causal framework.
Indeed, machine-learning algorithms have often extracted non-causal associations
between the intervention and the outcome, improper for decision-making
\cite{winkler2019association,badgeley2019deep,obermeyer2019dissecting}.
Machine learning studies in medicine often rely on an implicit causal
thinking, via a good understanding of the clinical settings.
A clear framework helps making
sure nothing falls through the cracks.

We have separated three steps important for causal validity: the choice
of study design, confounders, and estimators.
Regarding study design, major caveats arise from the time component,
where a poor choice of inclusion time easily brings in significant bias. Regarding choice of prediction
variables, forgetting some variables that explains both the treatment
allocation and the outcome leads to confounding bias, that however
remains small when these
variables capture weak links. Regarding choice of causal estimators,
preferring flexible models such as random forests reduces the bias, in
particular for doubly-robust estimators.
We have shown that all these three steps are equally important: paying no
attention to one of them leads to invalid estimates of treatment effect,
yet imperfect but plausible choices lead to small biases of the same
order of magnitude for all steps.
For instance, despite the emphasis often put on choice of confounders,
minor deviations from the expert's causal graph did not introduce
substantial bias (\nameref{apd:vibration_analysis_for_confounders}), no larger than a too
rigid choice of estimator.
To assert the validity of the analysis, we argue to relate as much as
possible the average effect to a reference target trial, even when the
goal is to capture the heterogeneity of the effect to individualize
decisions.
EHRs complement RCTs: RCTs cannot address all the
subpopulations and local practices
\cite{travers2007external,kennedy2015literature}. EHRs often cover many
individuals, with the diversity needed to model treatment
heterogeneity. The corresponding model can then inform better
decision-making \cite{prosperi2020causal}: a sub-population analysis  (as in Figure
\ref{fig:albumin_for_sepsis:cate_results}) can distill rules on which groups
of patients should receive a treatment. Beyond a sub-group perspective,
patient-specific estimates facilitate a personalized approach
to clinical decision-making \cite{kent2018personalized}.

Since the early 1980ies, researcher investigated the use of colloid fluids in
sepsis resuscitation due to their theoretical advantages. However, evidence has
long been conflicting. The debate was sparked anew when new synthetic colloid
solutions became available, but were later shown to have renal adverse effects
\cite{xu2014comparison}. As even large RCTs left unanswered questions,
researchers focused on meta-analyses. Here our analysis is in line with the
latest two meta-analyses \cite{xu2014comparison,li2020resuscitation}, as we
found no net benefit for resuscitation with albumin in septic patients overall,
but a possible slight benefit for patients with septic shock (see Fig.
\ref{fig:albumin_for_sepsis:cate_results}). While regular meta-analyses not
utilizing patient-level data are restricted in their sensitivity analyses, our
approach offers the benefit to investigate further potential effect modifiers
such as age, sex, or race.

Even without considering a specific intervention, anchoring
machine-learning models on
causal mechanisms can make them more robust to distributional shift \cite{scholkopf2021toward},
thus safer and fairer for clinical use
\cite{richens2020improving,plecko2022causal}.
Yet it is important to keep in mind that better prediction is not per se
a goal in healthcare.
Establishing strong predictors might be less important than identifying
moderately strong but modifiable risk factors as established in the Framingham
cohort \cite{brand1976multivariate}, or optimizing population-wide cost-effectiveness instead of individual treatment effect.

No sophisticated data-processing tool can safeguard against
invalid study design or a major missing confounder, loopholes that can
undermine decision-making systems. Our framework helps the investigator
ensure
causal validity by outlining the important steps and relating average effects to
RCTs. Causal grounding of individual predictions should reduce the social
disparities that they reinforce
\cite{rajkomar2018ensuring,mitra2022future,ehrmann2023making}, as these are driven by
historical decisions and not biological mechanisms. At the population
level, it leads to better public health decisions. For instance, going
back to cardio-vascular diseases, the stakes are to go beyond risk
scores and also account for responder status when prescribing prevention
drugs.

\section*{Availability of data and materials}

The datasets are available on PhysioNet (
\url{https://doi.org/10.13026/6mm1-ek67}). We used MIMIC-IV.v2.2 The code for
data preprocessing and analyses are available on github
\url{https://github.com/soda-inria/causal_ehr_mimic/}.
The project was run on a laptop running Ubuntu 22.04.2 LTS with the following hardware: CPU 12th Gen Intel(R) Core(TM) i7-1270P with 16 threads and 15 GB of RAM.


\section*{Authors contributions}

MD and TS designed the study, MD performed the analysis and wrote the manuscript.
TS, JA, CM, LAC, GV reviewed and edited the manuscript.

\section*{Acknowledgments}

We thank all the PhysioNet team for their encouragements and support. In
particular: Fredrik Willumsen Haug, João Matos, Luis Nakayama, Sicheng Hao, Alistair Johnson.

\nolinenumbers

\bibliography{references}

\clearpage

\section*{Supporting information}

\begin{figure}[!h]
  \begin{subfigure}{\linewidth}
    \centering
    \includegraphics[width=0.8\linewidth]{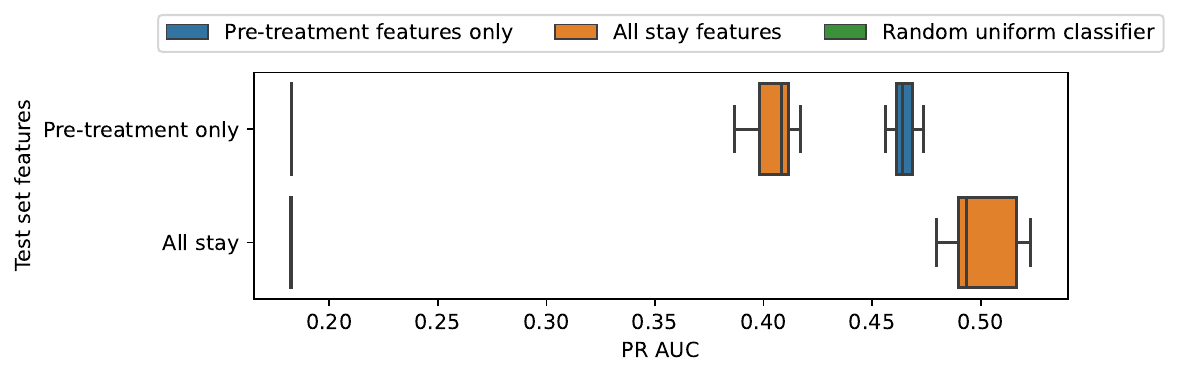}
    \caption{Area under the Precision-Recall curve (PR\_AUC)}\label{apd:fig:motivating_example_pr_auc}
  \end{subfigure}
  \hfill
  \begin{subfigure}{\linewidth}
    \centering
    \includegraphics[width=0.8\linewidth]{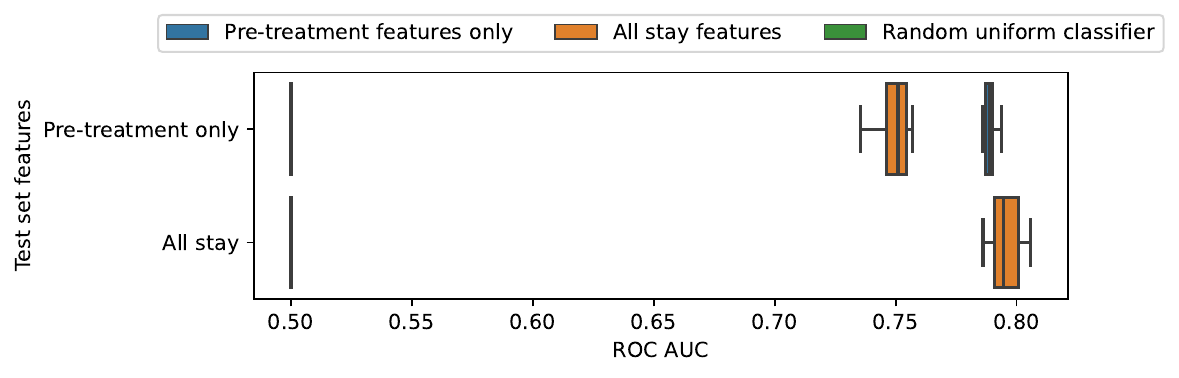}
    \caption{Area under the Receiving Operator Characteristic (ROC\_AUC)}\label{apd:fig:motivating_example_roc_auc}
  \end{subfigure}
  \caption{Failure to predict 28-day mortality from a model fitted on
    pre-treatment variables. The model is trained on the last features from
    the whole stay and tested on two validation sets: one with all stay
    features and one with last features before crystalloids administration
    (Pre-treatment only). The all-stay model performance markedly decreases in
    the pre-treatment only dataset.}\label{apd:fig:motivating_example}
\end{figure}

\paragraph*{S1 Fig.}
\label{apd:motivating_example}
{\bf Motivating example: Failure of predictive models to predict mortality
  from pretreatment variables.}
To illustrate how machine learning frameworks can fail to inform decision
making, we present a motivating example from MIMIC-IV. Using the same
population and covariates as in the main analysis (described in Table
\ref{apd:table:albumin_for_sepsis:table1_complete}), we train a predictive
model for 28-day mortality. We split the data into a training set (80\%) and a
test set (20\%). The training set uses the last measurements from the first 24
hours, whereas the validation set only uses the last measurements before the
administration of crystalloids. We split the train set into a train and a
validation set. We fit a HistGradientBoosting classifier
\footnote{\url{https://scikit-learn.org/stable/modules/ensemble.html\#histogram-based-gradient-boosting}}
on the train set and evaluate the performance on the validation set and on the
test set. We see good area under the Precision-recall curve (PR AUC) on the
validation set, but a deterioration of 10 points on the test set (Figure
\ref{apd:fig:motivating_example_pr_auc}). The same is seen in Figure
\ref{apd:fig:motivating_example_roc_auc} when measuring performances with Area
Under the Curve of the Receiving Operator Characteristic (ROC AUC). On the
contrary, a model trained on pre-treatment features yields competitive
performances. This failure illustrates well the shortcuts on which predictive
models could rely to make predictions. A clinically useful predictive model
should support decision-making --in this case, addition of albumin to
crystalloids-- rather than maximizing predictive performance. In this example,
causal thinking would have helped to identify the bias introduced by
post-treatment features. In fact, these features should not be included in a
causal analysis since they are post-treatment colliders.

This kind of error might sound naive to a clinical expert but relying on
shortcuts --some of them being post-treatment variables-- is a common error.
Here, we detail some real use cases where machine learning fail in providing
useful predictions for decision-making. \cite{badgeley2019deep} use deep
learning to predict hip fracture using confounding patient and healthcare
variables. An example of such covariates shown by the authors is the triage of
patients before imaging that results in the model trying to predict the image
acquisition machine and rely on it to predict hip fracture.
\cite{obermeyer2019dissecting} describe the use of algorithm in US extra-care
programs. By equating care needs with previous care costs (in a pure
predictive fashion), the algorithm falsely conclude that Black patients are
healthier than equally white patients, since they do less money is spent on
them for a given level of need. Beyond Machine Learning, we also spotted the
inclusion of post-treatment variables in the development of the recent SCORE2
cardio-vascular risk score \cite{score22021score2}: \emph{Our risk models
  might have underestimated CVD risk be- cause data used to estimate multipliers
  were likely to include some people already on CVD prevention therapies (e.g.
  statins or anti- hypertensive medication}. This score might be used to inform
on the initiation of statins for primary prevention. But, relying on
post-treatment, it might under-discover patients who would benefit from
statins at screening time.

\clearpage

\paragraph*{S2 Fig.}
\label{apd:immortal_time_bias}
{\bf Immortal time bias illustration.}

\begin{figure}[!h]
  \includegraphics[width=0.7\linewidth]{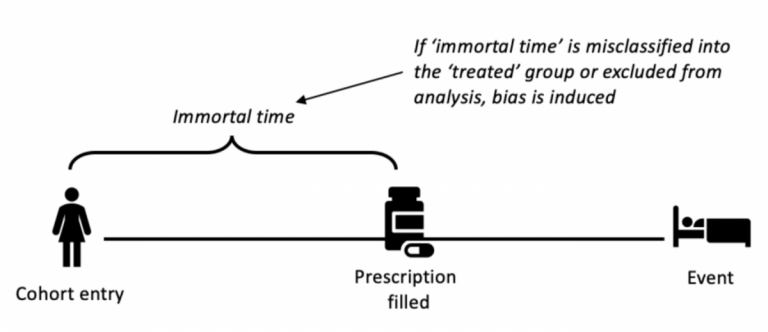}
  \caption{Poor experimental design can introduce Immortal time bias, which
    leads to a treated group with falsely longer longevity
    \cite{lee2020immortaltimebias}.}\label{apd:fig:immortal_time_bias}
\end{figure}

Figure \ref{apd:fig:immortal_time_bias} illustrates the immortal time bias.
This time bias is a major pitfall in the retrospective evaluation of
screening programs  \cite{bretthauer2013principles}.

\clearpage

\begin{figure}[h!]
  \centering
  \includegraphics[width=\linewidth]{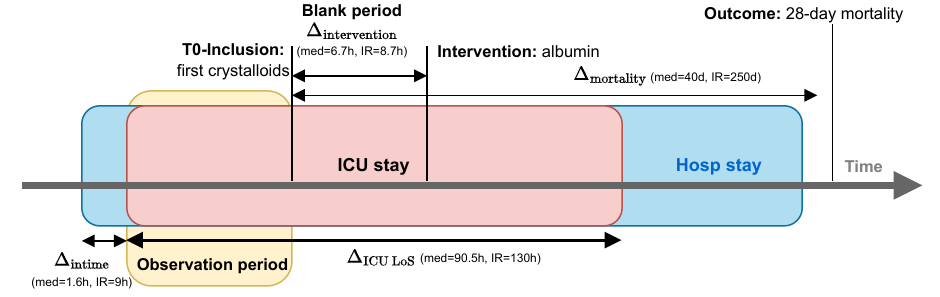}
  \caption{Defining the inclusion event, the starting time T0 for follow-up, the
    intervention's assignment time and the observation window for confounders is
    crucial to avoid time and selection biases. In our study, the gap
    between the intervention and the inclusion is small
    compared to the occurrence of the outcome to limit immortal time bias: 6.7 hours vs 40
    days for mortality.\label{fig:cohort_timeline_albumin}}
\end{figure}

\paragraph*{S3 Fig.}
\label{apd:graphical_timeline}
{\bf Graphical timeline.}
Drawing a graphical timeline as the one in Figure
\ref{fig:cohort_timeline_albumin} during the study design helps to detect and
prevent time-related biases.

\clearpage

\paragraph*{S4 Fig.}
\label{apd:causal_variables}
{\bf Types of causal variables.}

\begin{figure}[h!]
  \begin{minipage}[t]{0.32\linewidth}
    \centering
    \includegraphics[width=0.72\linewidth]{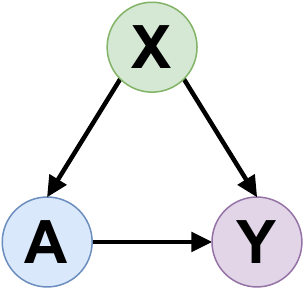}
    \small\sffamily Confounder
  \end{minipage}
  \hfill
  \begin{minipage}[t]{0.32\linewidth}
    \centering
    \includegraphics[width=0.72\linewidth]{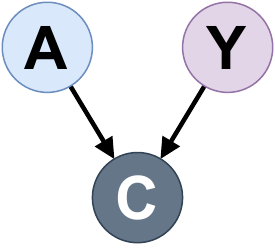}
    \small\sffamily Collider
  \end{minipage}
  \hfill
  \begin{minipage}[t]{0.32\linewidth}
    \centering
    \includegraphics[width=0.72\linewidth]{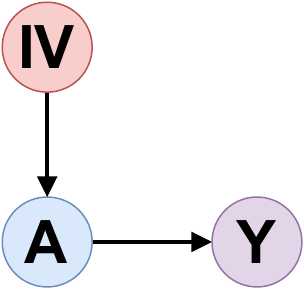}
    \small\sffamily Instrumental \\variable
  \end{minipage}
  \vfill
  \begin{minipage}[t]{0.4\linewidth}
    \centering
    \includegraphics[width=.9\linewidth]{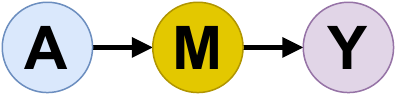}%
    \\
    \small\sffamily Mediator
  \end{minipage}
  \hfill
  \begin{minipage}[t]{0.4\linewidth}
    \centering
    \includegraphics[width=\linewidth]{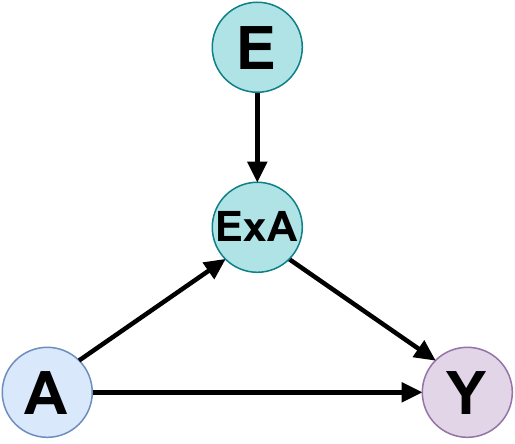}%
    \small\sffamily Effect modifier \scriptsize\\Represented
    following \cite{attia2022proposal}
  \end{minipage}
  \caption{The five categories of causal variables needed for our framework: A:
    Treatment, X: Confounder, IV: Instrumental variable, M: mediator, Y:
    Outcome, C: Collider, E: Effect modifier.}\label{fig:causal_variables}
\end{figure}

Figure \ref{fig:causal_variables} illustrates the different types of causal variables.

\clearpage

\paragraph*{S1 Appendix.}
\label{apd:causal_assumptions}
{\bf Assumptions: what is needed for causal inference from observational studies.}

The following four assumptions, referred as strong ignorability, are
needed to assure identifiability of the causal estimands with observational
data with most causal-inference methods \cite{rubin2005causal}, in
particular these we use:

\begin{assumption}[Unconfoundedness]\label{assumption:ignorability}
  \begin{equation}\label{eq:ignorability}
    \{Y(0), Y(1) \} \indep A | X
  \end{equation}
  This condition --also called ignorability-- is equivalent to the conditional
  independence on the propensity score $e(X)=\mathbb P(A=1|X)$ \cite{rosenbaum1983central}: $\{Y(0), Y(1) \}\indep  A | e(X)$.
\end{assumption}

Unconfoundedness is a strong assumption that might be violated in practice. The
existence of residual bias trough unobserved confounders can be mitigated with
different strategies. The \textit{omitted variable bias} framework encourages
sensitivity analyses allowing to derive bounds on the causal estimate by making
assumptions on the strength of association of the omitted variable with both the
treatment and the outcome. We refer to \cite{cinelli2020making} for a clear
introduction under linear assumption and to \cite{chernozhukov2022long} for an
extension to general non-linear settings. In case of strong unobserved
confounders for which proxy variables can be measured, \textit{proximal
  inference} can be used to obtain identifiability
\cite{tchetgen2024introduction}. These methods require expert knowledge to
classify the proxy between treatment and outcome proxy, after which a two-stage
regression is run to recover the causal effect. Lastly, natural experiments, when
available, should be exploited to estimate causal effects without the need of
unconfoundedness. Instrumental variable methods exploit randomness influencing
the treatment but unrelated to the outcome to simulate a randomized experiment
\cite[chapter 9]{wager2020stats}. Regression discontinuity designs leverage
discontinuous treatment assignment mechanisms with the assumption of a
continuous outcome \cite[chapter 5]{wager2020stats}.

\begin{assumption}[Overlap, also known as Positivity]\label{assumption:overlap}
  \begin{equation}\label{eq:overlap}
    \eta < e(x) < 1 - \eta \quad \forall x \in \mathcal{X} \text{ and some }   \eta > 0
  \end{equation}
  The treatment is not perfectly predictable. Or in other words, every
  patient has a chance to be treated and not to be treated. For a given set of
  covariates, we need examples of both to recover the ATE.
\end{assumption}

As noted by \cite{damour2020overlap}, the choice of covariates $X$ can
be viewed as a trade-off between these two central assumptions. A bigger
covariate set generally reinforces the ignorability assumption. In the
contrary, overlap can be weakened by large $\mathcal{X}$ because of the
potential inclusion of instrumental variables: variables only linked to the treatment which
could lead to arbitrarily small propensity scores.


\begin{assumption}[Consistency]\label{assumption:consistency} The observed
  outcome is the potential outcome of the assigned treatment:
  \begin{equation}\label{eq:consistancy}
    Y = A \, Y(1) + (1-A) \, Y(0)
  \end{equation}
  Here, we assume that the intervention $A$ has been well defined. This
  assumption focuses on the design of the experiment. It clearly states the link
  between the observed outcome and the potential outcomes through the
  intervention \cite{hernan2020causal}.
\end{assumption}

\begin{assumption}[Generalization]\label{assumption:generalization} The training
  data on which we build the estimator and the test data on which we make the
  estimation are drawn from the same distribution, also known as
  the ``no covariate shift'' assumption \cite{jesson2020identifying}.
\end{assumption}
\clearpage

\paragraph*{S2 Appendix.}
\label{apd:causal_estimators}
{\bf Major causal-inference methods: When to use which estimator?}

\textbf{G-formula} also called conditional mean regression
\cite{wendling2018comparing}, g-computation \cite{robins1986role}, or
Q-model \cite{snowden2011implementation}. This approach is directly modeling
the outcome, also referred to as the response surface: $\mu_{(a)}(x)
  =\mathbb{E}\left(Y \mid A=a, \mathbf{X}=x\right)$

Using an outcome estimator to learn a model for the response surface $\hat
  \mu$ (eg. a linear model), the ATE estimator is an average over the n samples:
\begin{equation}
  \hat{\tau}_G(f) = \frac{1}{n} \sum_{i=1}^n \hat \mu(x_i, 1) - \hat \mu(x_i, 0) = \frac{1}{n} \sum_{i=1}^n \hat \mu_{(1)}(x_i) - \hat \mu_{(0)}(x_i)
\end{equation}

This estimator is unbiased if the model of the conditional response surface
$\hat \mu_{(a)}$ is well-specified. This approach assumes than $Y(a) =
  \mu_a(X) + \epsilon_a$ with $\mathbb E[\epsilon|X] = 0$. The main drawback is
the extrapolation of the learned outcome estimator from samples with similar
covariates X but different intervention A.\\

\textbf{Propensity Score Matching (PSM)} To avoid confounding bias, the
ignorability assumption \ref{assumption:ignorability}) requires to contrast
treated and control outcomes only between comparable patients with respect to
treatment allocation probabilities. A simple way to do this is to group
patients into bins, or subgroups, of similar confounders and contrast the two
population's outcomes by matching patients inside these bins
\cite{stuart2010matching}. However, the number of confounder bins grows
exponentially with the number of variables. \cite{rosenbaum1983central} proved
that matching patients on the individual probabilities to receive treatment
--propensity scores-- is sufficient to verify ignorability. PSM is a
conceptually simple method, but has delicate parameters to tune such as
choosing a model for the propensity score, deciding what is the maximum
distance between two potential matches (the caliper width), the number of
matches by sample, and matching with or without replacement. It also prunes data
not meeting the caliper width criteria, and suffers from high estimation
variance in highly-dimensional data where extreme propensity weights are common.
Finally, the simple bootstrap confidence intervals are not theoretically
grounded \cite{abadie2008failure} making PSM more difficult
to use for applied practitioners.\\

\textbf{Inverse Propensity Weighting (IPW)}

A simple alternative to propensity score matching is to weight the outcome by
the inverse of the propensity score \cite{austin2015moving}. It relies on a
similar idea as matching but automatically builds a balanced population by
reweighting the outcomes with the propensity score model $\hat{e}$ to estimate
the ATE:
\begin{equation}
  \hat \tau_{IPW}(\hat e) = \frac{1}{n} \sum_{i=1}^N \frac{A_i Y_i}{\hat e(X_i)} - \frac{(1-A_i)Y_i}{(1-\hat e(X_i))}
\end{equation}

This estimate is unbiased if $\hat e$ is well-specified. IPW suffers from high
variance if some weights are too close to 0 or 1. In high dimensional cases
where poor overlap between treated and control is common, one can clip extreme
weights to limit estimation instability.\\

\textbf{Doubly Robust Learning, DRL} also called Augmented Inverse
Probability Weighting (AIPW) \cite{robins1994estimation}.

The underlying idea of DRL is to combine the G-formula and IPW estimators to
protect against a mis-specification of one of them. It first requires to
estimate the two nuisance parameters: a model for the intervention $\hat{e}$
and a model for the outcome $f$. If one of the two nuisance is unbiased, the
following ATE estimator is as well:

$$\begin{aligned} \widehat{\tau}_{A I P W}=\frac{1}{n}
    \sum_{i=1}^{n}\left(\hat \mu_{(1)}\left(x_{i}\right)-\hat \mu_{(0)}\left(x_{i}\right)+a_{i}
    \frac{y_{i}-\hat \mu_{(1)}\left(x_{i}\right)}{\hat{e}\left(x_{i}\right)}-\left(1-a_{i}\right)
    \frac{y_{i}-\hat \mu_{(0)}\left(x_{i}\right)}{1-\hat{e}\left(x_{i}\right)}\right)
  \end{aligned}$$

Moreover, despite the need to estimate two models, this estimator is more
efficient in the sense that it converges quicker than single model estimators
\cite{wager2020stats}. For this propriety to hold, one need to fit and apply
the two nuisance models in a cross-fitting manner. This means that we split
the data into K folds. Then for each fold, we fit the nuisance models on the
K-1 complementary folds, and predict on the remaining fold.

To recover Conditional Treatment Effects from the AIPW estimator,
\cite{foster2019orthogonal} suggested to regress the Individual Treatment
Effect estimates from AIPW on potential sources of heterogeneity $X^{cate}$:
$\hat tau = \argmin_{\tau \in \Theta} (\hat \tau_{AIPW}(X) - \tau(X^{cate}))$
for $\Theta$ some class of model (eg. linear model).\\

\textbf{Double Machine Learning} \cite{chernozhukov2018double} also known
as the R-learner \cite{nie2021quasi}. It is based on the R-decomposition,
\cite{robinson1988root}, and the modeling of the conditional mean outcome,
$m(x)=\mathbb E[Y|X=x]$ and the propensity score, $e(x)=\mathbb E[A=1|X=x]$:
\begin{equation}\label{eq:r_decomposition}
  y_{i}-m\left(x_{i}\right)=\left(a_{i}-e\left(x_{i}\right)\right) \tau\left(x_{i}\right)+\varepsilon_{i} \quad \text{with} \; \varepsilon_{i}=y_{i}-\varepsilon\left[4_{i} \mid x_{i}, a_{i}\right]
\end{equation}
Note that we can impose that the conditional treatment effect $\tau(x)$ only
relies on a subset of the features, $x^{cate}$ on which we want to study
treatment heterogeneity.

From this decomposition, we can derive an estimation of the ATE $\tau$, where
the right hand-side term is the empirical R-Loss:
\begin{align}\label{eq:r_loss}
  \hat{\tau}(\cdot)=\operatorname{argmin}_{\tau}\left\{\frac{1}{n} \sum_{i=1}^{n}\left(\left(y_{i}-m\left(x_{i}\right)\right)-\left(a_{i}-e(x_{i})\right) \tau\left(x^{cate}_{i}\right)\right)^{2}\right\}
\end{align}

The full procedure for R-learning is first to fit the nuisances: $\hat m$ and
$\hat e$. Then, minimize the estimated R-loss eq.\ref{eq:r_loss}, where
the oracle nuisances $(e, m)$ have been replaced by their estimated
counterparts $(\hat e, \hat m)$. Minimization can be done by regressing
the outcome residuals weighted by the treatment residuals. Finally, get
the ATE by averaging conditional treatment effect $\tau(x^{cate})$ over
the population.

This estimator has also the doubly robust proprieties described for AIPW. it
should have less variance than AIPW since it does not use the propensity score
in the denominator.
\clearpage

\paragraph*{S3 Appendix.}
\label{apd:statistical_considerations}
{\bf Statistical considerations when implementing
  estimation.}

\textbf{Counterfactual prediction lacks off-the-shelf cross-fitting estimators}

Doubly robust methods use cross-fit estimation of the nuisance parameters,
which is not available off-the-shelf for IPW and T-Learner estimators. For
reproducibility purposes, we did not reimplement internal cross-fitting for
treatment or outcome estimators. However, when flexible models such as
random forests are used, a fairer comparison between single and double
robust methods should use cross-fitting for both. This lack in the
scikit-learn API \cite{pedregosa2011scikit} reflects different needs
between purely predictive machine learning focused on generalization
performances and counterfactual prediction aiming at unbiased inference on
the input data.

\textbf{Good practices for imputation not implemented in EconML}

Good practices in machine learning recommend to input distinctly each fold
when performing cross-fitting
\footnote{\url{https://scikit-learn.org/stable/modules/compose.html\#combining-estimators}}.
However, EconML estimators test for missing data at instantiation
preventing the use of scikit-learn imputation pipelines. We thus have been
forced to transform the full dataset before feeding it to causal estimators.
An issue mentioning the problem has been filed, so we can hope that future
versions of the package will comply with best practices. \footnote{\url{https://github.com/py-why/EconML/issues/664}}

\textbf{Bootstrap may not yield the most efficient confidence intervals}

To ensure a fair comparison between causal estimators, we always used
bootstrap estimates for confidence intervals. However, closed form
confidence intervals are available for some estimators -- see \cite{wager2020stats}
for IPW and AIPW (DRLeaner) variance estimations. These formulas exploit the
estimator properties, thus tend to have smaller confidence intervals. On the
other hand, they usually do not include the variance of the outcome and
treatment estimators, which is naturally dealt with in bootstrapped confidence
intervals. Closed form confidence intervals are rarely implemented in any of the
packages as Dowhy for the IPW estimator, or in EconML for AIPW.

Bootstrap was particularly costly to run for the EconML doubly robust
estimators (AIPW and Double ML), especially when combined with random forest nuisance
estimators (from 10 to 47 min depending on the aggregation choice and the
estimator). See Table \ref{apd:table:compute_times} for details.

\begin{table}[]
  \centering\small
  \begin{tabular}{llrll}
\toprule
{} &                    estimation\_method &  compute\_time &   outcome\_model & event\_aggregations \\
\midrule
2  &                            LinearDML &   1127.977827 &         Forests &  ['first', 'last'] \\
3  &   backdoor.propensity\_score\_matching &    199.765587 &         Forests &  ['first', 'last'] \\
4  &  backdoor.propensity\_score\_weighting &     86.149872 &         Forests &  ['first', 'last'] \\
5  &                             TLearner &    284.066786 &         Forests &  ['first', 'last'] \\
6  &                      LinearDRLearner &   2855.403709 &         Forests &  ['first', 'last'] \\
7  &                            LinearDML &     49.911035 &  Regularized LR &  ['first', 'last'] \\
8  &   backdoor.propensity\_score\_matching &    127.929910 &  Regularized LR &  ['first', 'last'] \\
9  &  backdoor.propensity\_score\_weighting &      6.407206 &  Regularized LR &  ['first', 'last'] \\
10 &                             TLearner &      6.843931 &  Regularized LR &  ['first', 'last'] \\
11 &                      LinearDRLearner &     80.747301 &  Regularized LR &  ['first', 'last'] \\
\bottomrule
\end{tabular}

  \\
  \caption{Compute times for the different estimation methods with 50 bootstrap replicates.}\label{apd:table:compute_times}
\end{table}
\clearpage

\paragraph*{S4 Appendix.}
\label{apd:packages}
{\bf Packages for causal estimation in the python ecosystem.}
We searched for causal inference packages in the python ecosystem. The focus
was on the identification methods. Important features were ease of
installation, sklearn estimator support, sklearn pipeline support, doubly
robust estimators, confidence interval computation, honest splitting
(cross-validation), Targeted Maximum Likelihood Estimation. These criteria are
summarized in \ref{apd:table:causal_packages}. We finally chose EconML despite
lacking \texttt{sklearn.\_BaseImputer} support through the
\texttt{sklearn.Pipeline} object as well as a TMLE implementation.

The zEpid package is primarily intended for epidemiologists. It is well documented
and provides pedagogical tutorials. It does not support sklearn estimators,
pipelines and honest splitting.

EconML \cite{battocchi2019econml} implements almost all estimators except propensity score methods. Despite
focusing on Conditional Average Treatment Effect, it provides all. One downside
is the lack of support for scikit-learn pipelines with missing value imputers.
This opens the door to information leakage when imputing data before splitting
into train/test folds.

Dowhy \cite{sharma2020dowhy} focuses on graphical models and relies on EconML for most of the causal
inference methods (identifications) and estimators. Despite, being interesting
for complex inference --such as mediation analysis or instrumental variables--,
we considered that it added an unnecessary layer of complexity for our use case
where a backdoor criterion is the most standard adjustment methodology.

Causalml implements all methods, but has a lot of package dependencies
which makes it hard to install.

\begin{table}[h!]
  \resizebox{\textwidth}{!}{%
    \begin{tabular}{|l|l|l|l|l|l|l|l|l|}
      \hline
      \multicolumn{1}{|c|}{\textbf{Packages}}                                                           &
      \multicolumn{1}{c|}{\textbf{\begin{tabular}[c]{@{}c@{}}Simple \\ installation\end{tabular}}}      &
      \multicolumn{1}{c|}{\textbf{\begin{tabular}[c]{@{}c@{}}Confidence \\ Intervals\end{tabular}}}     &
      \textbf{\begin{tabular}[c]{@{}l@{}}sklearn \\ estimator\end{tabular}}                             &
      \textbf{\begin{tabular}[c]{@{}l@{}}sklearn \\ pipeline\end{tabular}}                              &
      \multicolumn{1}{c|}{\textbf{\begin{tabular}[c]{@{}c@{}}Propensity \\ estimators\end{tabular}}}    &
      \multicolumn{1}{c|}{\textbf{\begin{tabular}[c]{@{}c@{}}Doubly Robust \\ estimators\end{tabular}}} &
      \multicolumn{1}{c|}{\textbf{\begin{tabular}[c]{@{}c@{}}TMLE\\ estimator\end{tabular}}}            &
      \multicolumn{1}{c|}{\textbf{\begin{tabular}[c]{@{}c@{}}Honest splitting \\ (cross validation)\end{tabular}}} \\ \hline
      \textbf{\href{https://github.com/py-why/dowhy}{dowhy}}                                            &
      \ding{51}                                                                                         &
      \ding{51}                                                                                         &
      \ding{51}                                                                                         &
      \ding{51}                                                                                         &
      \ding{51}                                                                                         &
      \ding{55}                                                                                         &
      \ding{55}                                                                                         &
      \ding{55}                                                                                                    \\ \hline
      \textbf{\href{https://github.com/py-why/EconML}{EconML}}                                          &
      \ding{51}                                                                                         &
      \ding{51}                                                                                         &
      \ding{51}                                                                                         &
      \begin{tabular}[c]{@{}l@{}}Yes except\\ for imputers\end{tabular}                                 &
      \ding{55}                                                                                         &
      \ding{51}                                                                                         &
      \ding{55}                                                                                         &
      \begin{tabular}[c]{@{}l@{}}Only for doubly \\ robust estimators\end{tabular}                                 \\ \hline
      \textbf{\href{https://github.com/pzivich/zEpid}{zEpid}}                                           &
      \ding{51}                                                                                         &
      \ding{51}                                                                                         &
      \ding{55}                                                                                         &
      \ding{55}                                                                                         &
      \ding{51}                                                                                         &
      \ding{51}                                                                                         &
      \ding{51}                                                                                         &
      Only for TMLE                                                                                                \\ \hline
      \textbf{\href{https://github.com/uber/causalml}{causalml}}                                        &
      \ding{55}                                                                                         &
      \ding{51}                                                                                         &
      \ding{51}                                                                                         &
      \ding{51}                                                                                         &
      \ding{51}                                                                                         &
      \ding{51}                                                                                         &
      \ding{51}                                                                                         &
      \begin{tabular}[c]{@{}l@{}}Only for doubly \\ robust estimators\end{tabular}                                 \\ \hline
    \end{tabular}%
  }\caption{Selection criteria for causal python packages}\label{apd:table:causal_packages}
\end{table}

\paragraph*{S5 Appendix.}
\label{apd:hyper_parameter_search}
{\bf Hyper-parameter search for the nuisance models.}
\begin{figure}[!b]
  \begin{minipage}{.38\linewidth}
    \caption{Hyper-parameter search procedure.}\label{apd:fig:hyper_parameter_search}
  \end{minipage}%
  \hfill
  \begin{minipage}{.6\linewidth}
    \includegraphics[width=\linewidth]{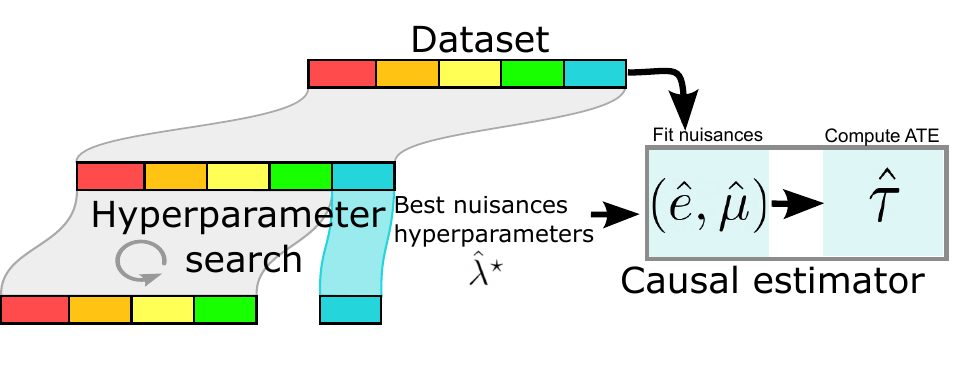}
  \end{minipage}%
\end{figure}
We followed a two-step procedure to train the nuisance models (eg. ($\hat e,
  \hat \mu$) for the AIPW causal estimator), taking inspiration from the
computationally cheap procedure from
\cite[section~3.3]{bouthillier2021accounting}. First, for each nuisance
model, we fit a random parameter search with 5-fold cross validation and 10
iterations on the full dataset. Each iteration fit a model with a random
combination of parameters in a predefined grid, then evaluate the
performance by cross-validation. The best hyper-parameters $\hat
  \lambda^{\star}$ are selected as the ones reaching the minimal score across
all iterations. Then, we feed this parameters to the causal estimator. The
single robust estimators (matching, IPW and TLearner) refit the
corresponding estimator only once on the full dataset, then estimate the
ATE. The doubly-robust estimators use a cross-fitting procedure (K=5) to fit
the nuisances then estimate the ATE. Figure
\ref{apd:fig:hyper_parameter_search} illustrates the procedure and Table
\ref{apd:table:hyper_parameter_search} details the hyper-parameters grid for
the random search.

\begin{table}[]
  \resizebox{\textwidth}{!}{%
    \begin{tabular}{llll}
    \toprule
    {}             & estimator              & nuisance  & Grid                                                                       \\
    Estimator type &                        &           &                                                                            \\
    \midrule
    Linear         & LogisticRegression     & treatment & \{'C': logspace(-3, 2, 10)\}                                               \\
    Linear         & Ridge                  & outcome   & \{'alpha': logspace(-3, 2, 10)\}                                           \\
    Forest         & RandomForestClassifier & treatment & \{'n\_estimators': ['10', '100', '200'], 'max\_depth': ['3', '10', '50']\} \\
    Forest         & RandomForestRegressor  & outcome   & \{'n\_estimators': ['10', '100', '200'], 'max\_depth': ['3', '10', '50']\} \\
    \bottomrule
\end{tabular}

  }\\
  \caption{Hyper-parameter grid used during random search
    optimization.}\label{apd:table:hyper_parameter_search}
\end{table}
\clearpage

\paragraph*{S5 Fig.}
\label{apd:selection_flowchart}
{\bf Selection flowchart.}

\begin{figure}[!htb]
  \centering
  \includegraphics[width=0.7\linewidth]{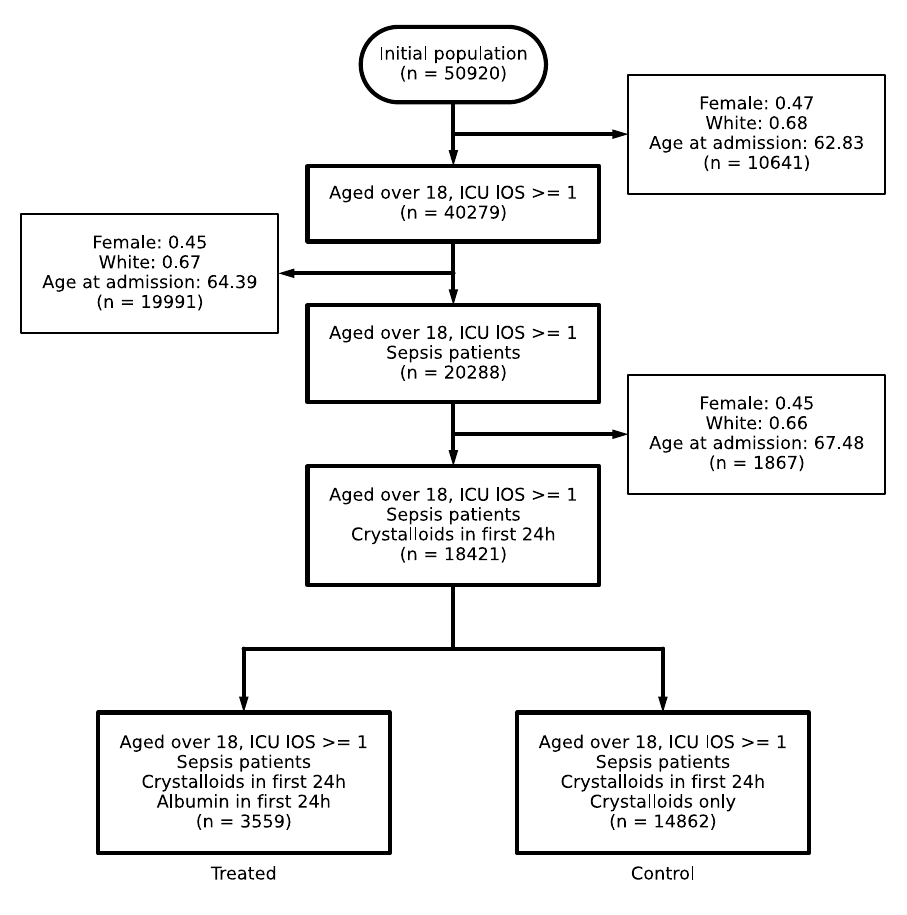}
  \caption{Selection flowchart on MIMIC-IV for the emulated trial.}\label{fig:selection_flowchart}
\end{figure}
\clearpage

\paragraph*{S1 Table.}
\label{apd:albumin_for_sepsis:table1_complete}
{\bf Complete description of the confounders for the main analysis.}
\begin{table}[h!]
  \resizebox{\textwidth}{!}{%
    \begin{tabular}{llllll}
\toprule
{} & Missing &        Overall & Cristalloids only & Cristalloids + Albumin & P-Value \\
                                                  &         &                &                   &                        &         \\
\midrule
n                                                 &         &          18421 &             14862 &                   3559 &         \\
Glycopeptide, n (\%)                               &         &    9492 (51.5) &       7650 (51.5) &            1842 (51.8) &         \\
Beta-lactams, n (\%)                               &         &    5761 (31.3) &       5271 (35.5) &             490 (13.8) &         \\
Carbapenems, n (\%)                                &         &      727 (3.9) &         636 (4.3) &               91 (2.6) &         \\
Aminoglycosides, n (\%)                            &         &      314 (1.7) &         290 (2.0) &               24 (0.7) &         \\
suspected\_infection\_blood, n (\%)                  &         &      170 (0.9) &         149 (1.0) &               21 (0.6) &         \\
RRT, n (\%)                                        &         &      229 (1.2) &         205 (1.4) &               24 (0.7) &         \\
ventilation, n (\%)                                &         &   16376 (88.9) &      12931 (87.0) &            3445 (96.8) &         \\
vasopressors, n (\%)                               &         &    9058 (49.2) &       6204 (41.7) &            2854 (80.2) &         \\
Female, n (\%)                                     &         &    7653 (41.5) &       6322 (42.5) &            1331 (37.4) &         \\
White, n (\%)                                      &         &   12366 (67.1) &       9808 (66.0) &            2558 (71.9) &         \\
Emergency admission, n (\%)                        &         &    9605 (52.1) &       8512 (57.3) &            1093 (30.7) &         \\
Insurance, Medicare, n (\%)                        &         &    9727 (52.8) &       7958 (53.5) &            1769 (49.7) &         \\
myocardial\_infarct, n (\%)                         &         &    3135 (17.0) &       2492 (16.8) &             643 (18.1) &         \\
malignant\_cancer, n (\%)                           &         &    2465 (13.4) &       2128 (14.3) &              337 (9.5) &         \\
diabetes\_with\_cc, n (\%)                           &         &     1633 (8.9) &        1362 (9.2) &              271 (7.6) &         \\
diabetes\_without\_cc, n (\%)                        &         &    4369 (23.7) &       3532 (23.8) &             837 (23.5) &         \\
metastatic\_solid\_tumor, n (\%)                     &         &     1127 (6.1) &        1016 (6.8) &              111 (3.1) &         \\
severe\_liver\_disease, n (\%)                       &         &     1289 (7.0) &         880 (5.9) &             409 (11.5) &         \\
renal\_disease, n (\%)                              &         &    3765 (20.4) &       3159 (21.3) &             606 (17.0) &         \\
aki\_stage\_0.0, n (\%)                              &         &    7368 (40.0) &       6284 (42.3) &            1084 (30.5) &         \\
aki\_stage\_1.0, n (\%)                              &         &    4019 (21.8) &       3222 (21.7) &             797 (22.4) &         \\
aki\_stage\_2.0, n (\%)                              &         &    6087 (33.0) &       4605 (31.0) &            1482 (41.6) &         \\
aki\_stage\_3.0, n (\%)                              &         &      947 (5.1) &         751 (5.1) &              196 (5.5) &         \\
SOFA, mean (SD)                                   &       0 &      6.0 (3.5) &         5.7 (3.4) &              6.9 (3.6) &  <0.001 \\
SAPSII, mean (SD)                                 &       0 &    40.3 (14.1) &       39.8 (14.1) &            42.8 (13.6) &  <0.001 \\
Weight, mean (SD)                                 &      97 &    83.3 (23.7) &       82.5 (24.2) &            86.4 (21.2) &  <0.001 \\
temperature, mean (SD)                            &     966 &     36.9 (0.6) &        36.9 (0.6) &             36.8 (0.6) &  <0.001 \\
mbp, mean (SD)                                    &       0 &    75.6 (10.2) &       76.3 (10.7) &             72.4 (7.2) &  <0.001 \\
resp\_rate, mean (SD)                              &       9 &     19.3 (4.3) &        19.6 (4.4) &             18.0 (3.8) &  <0.001 \\
heart\_rate, mean (SD)                             &       0 &    86.2 (16.3) &       86.2 (16.8) &            86.5 (14.3) &   0.197 \\
spo2, mean (SD)                                   &       4 &     97.4 (2.2) &        97.3 (2.3) &             98.0 (2.1) &  <0.001 \\
lactate, mean (SD)                                &    4616 &      3.0 (2.5) &         2.8 (2.4) &              3.7 (2.6) &  <0.001 \\
urineoutput, mean (SD)                            &     301 &    24.0 (52.7) &       24.7 (58.2) &            21.1 (16.6) &  <0.001 \\
admission\_age, mean (SD)                          &       0 &    66.3 (16.2) &       66.1 (16.8) &            67.3 (13.1) &  <0.001 \\
delta mortality to inclusion, mean (SD)           &   11121 &  316.9 (640.2) &     309.6 (628.8) &          365.0 (708.9) &   0.022 \\
delta intervention to inclusion, mean (SD)        &   14862 &      0.3 (0.2) &         nan (nan) &              0.3 (0.2) &     nan \\
delta inclusion to intime, mean (SD)              &       0 &      0.1 (0.2) &         0.1 (0.2) &              0.1 (0.1) &   0.041 \\
delta ICU intime to hospital admission, mean (SD) &       0 &      1.1 (3.7) &         1.0 (3.7) &              1.6 (3.4) &  <0.001 \\
los\_hospital, mean (SD)                           &       0 &    12.6 (12.5) &       12.6 (12.5) &            12.9 (12.4) &   0.189 \\
los\_icu, mean (SD)                                &       0 &      5.5 (6.7) &         5.5 (6.5) &              5.5 (7.2) &   0.605 \\
\bottomrule
\end{tabular}

  }\\
  \caption{Characteristics of the trial population measured on the first 24
    hours of ICU stay. \\
    Risk scores (AKI, SOFA, SAPSII) and lactates have been summarized as the
    maximum value during the 24 hour period for each stay. Total cumulative urine output has
    been computed. Other variables have been aggregated by taking mean during
    the 24 hour period.}\label{apd:table:albumin_for_sepsis:table1_complete}
\end{table}
\clearpage

\paragraph*{S6 Fig.}
\label{apd:causal_diagram_albumin}
{\bf Directed Acyclic Graph.}

The expert DAG in figure depicts the known causal links between these
variables.

\begin{figure}[h!]
  \centering%
  \includegraphics[width=\linewidth]{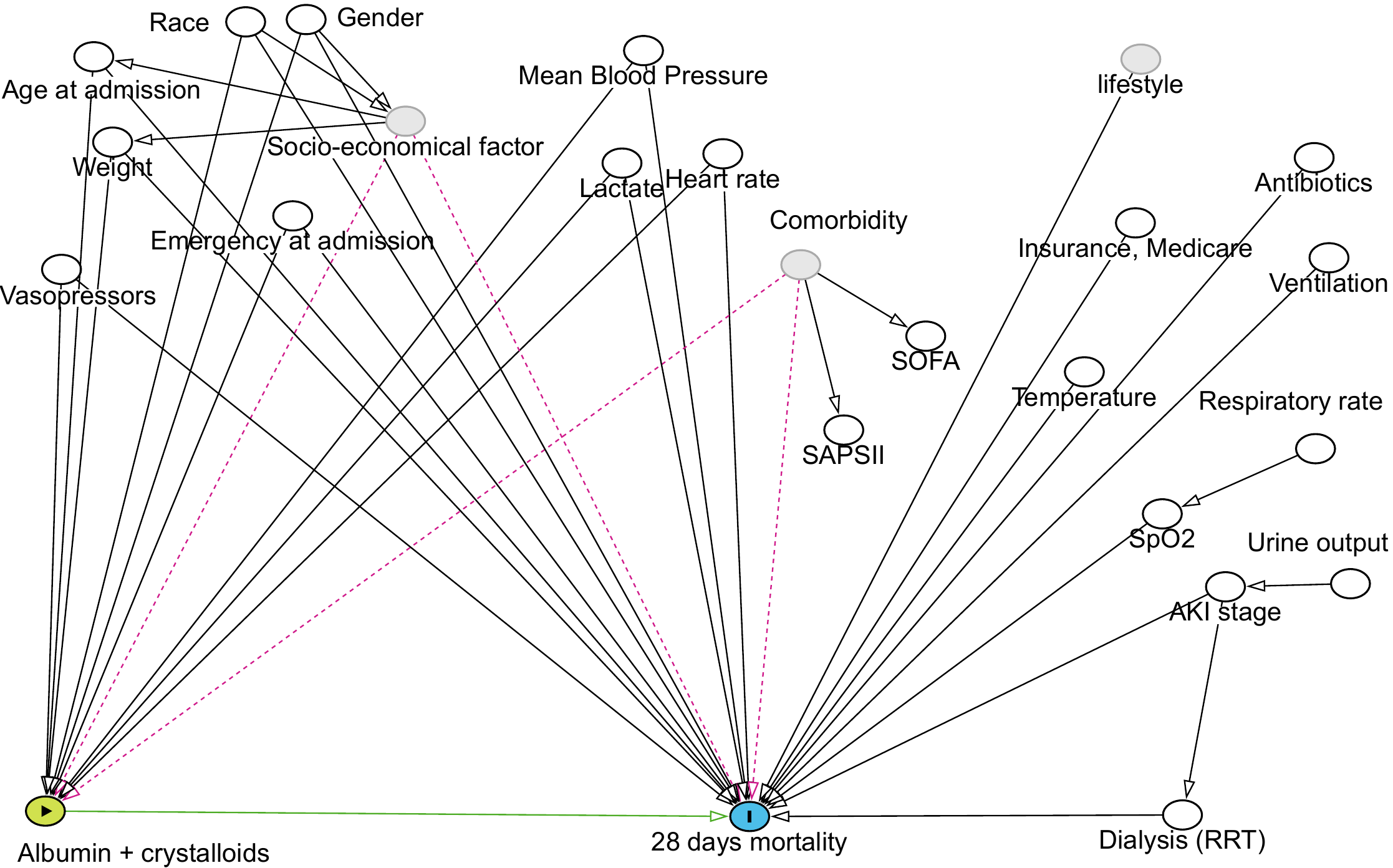}
  \caption{\textbf{Causal graph for the Albumin vs crystalloids emulated
      trial} -- The
    green arrow indicates the effect studied. Black arrows show causal links
    known to medical expertise. Dotted red arrows highlight confounders not directly
    observed. For readability, we draw only the most important edges from an
    expert point of view. All white nodes correspond to variables included in
    our study.}\label{fig:causal_diagram_albumin}
\end{figure}
\clearpage

\paragraph*{S7 Fig.}
\label{apd:detailed_results}
{\bf Complete results for the main analysis.}

\begin{figure}[h!]
  \centering
  \includegraphics[width=\linewidth]{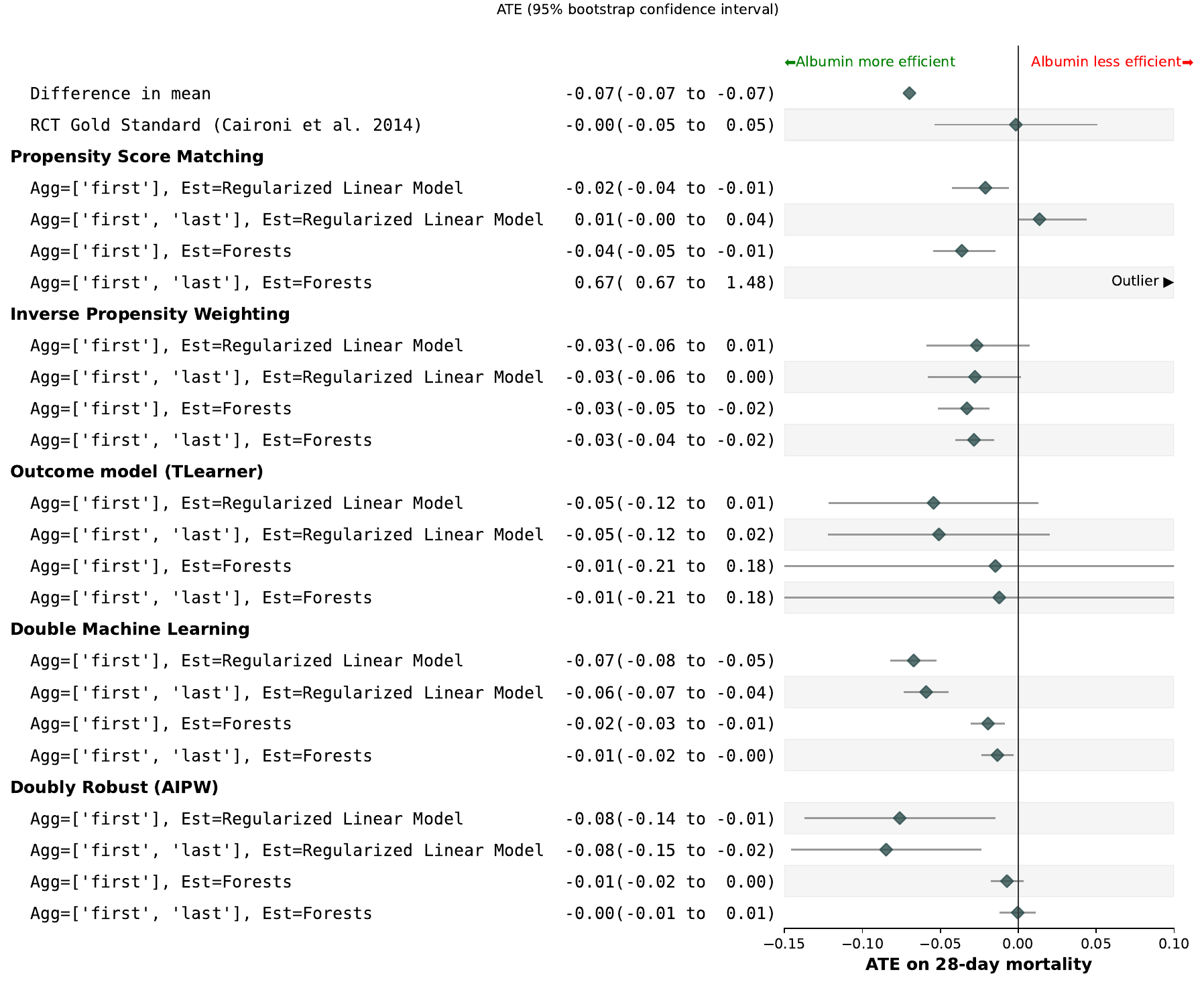}
  \caption{Full sensitivity analysis: The estimators with forest nuisances
    point to no effect for almost every causal estimator consistently with the
    RCT gold standard. Only matching with forest yields an unconvincingly high
    estimate. Linear nuisance used with doubly robust methods suggest a
    reduced mortality risk for albumin. The choices of aggregation only
    marginally modify the results expect for propensity score matching. The
    green diamonds depict the mean effect and the bar are the 95\% confidence
    intervals obtained by 50 bootstrap
    repetitions.}\label{apd:fig:detailed_results}
\end{figure}

Compared to Figure \ref{fig:vibration_analysis}, we also report in
Figure \ref{apd:fig:detailed_results} the estimates for Causal forest
estimators and other choices of feature aggregation (first and last).
\clearpage

\paragraph*{S8 Fig.}
\label{apd:detailed_results_itb}
{\bf Complete results for the Immortal time bias.}

\begin{figure}[h!]
  \centering
  \includegraphics[width=\linewidth]{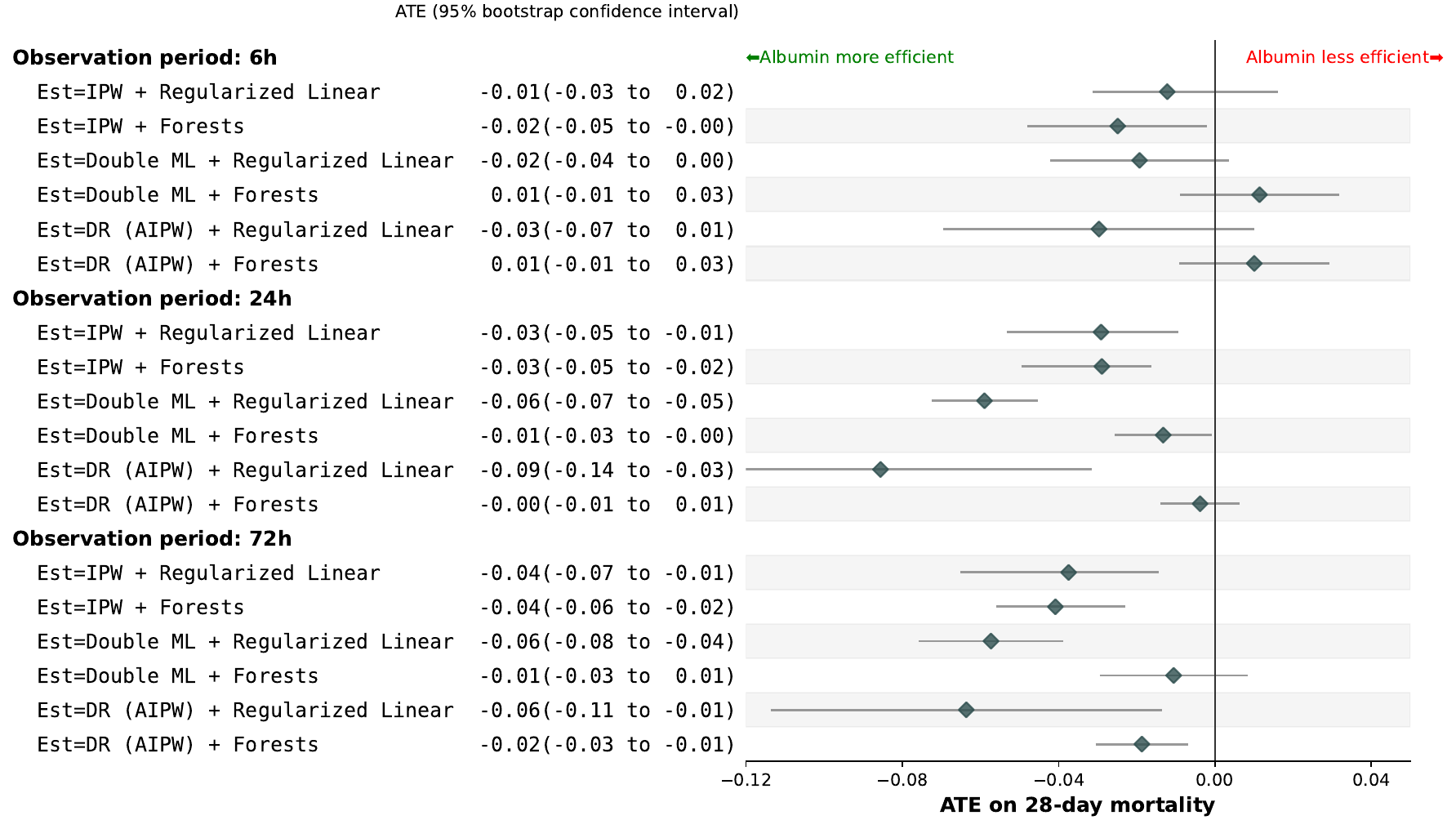}
  \caption{Sensitivity analysis for immortal time bias: Every choice of
    estimates show an improvement of the albumin treatment when increasing the
    observation period, thus increasing the blank period between inclusion and
    administration of albumin. Aggregation was concatenation of first and last
    features. The green diamonds depict the mean effect and the bar are the
    95\% confidence intervals obtained by 50 bootstrap
    repetitions.}\label{apd:fig:detailed_results_itb}
\end{figure}

Compared to Figure \ref{fig:vibration:itb}, we also report in Figure
\ref{apd:fig:detailed_results_itb} the estimates for Double Machine Learning,
Inverse Propensity Weighting for both Random Forest and Ridge Regression.
Feature aggregation was concatenation of first and last for all estimates.

\clearpage

\paragraph*{S6 Appendix.}
\label{apd:vibration_analysis_for_confounders}
{\bf Deviating from expert ignorability -- Impact of smaller confounders sets.}

We conducted a dedicated vibration analysis on the different choices of
confounders. We created three confounder subsets in addition to all
confounders (24 variables): all confounders without antibiotics
(Glycopeptides, Beta-lactams, Carbapenems, Aminoglycosides), all confounders
without any measurement (weight, lactate, heart rate,
spo2, mbp, urine output, temperature, AKI stage, SAPSII, respiratory rate, SOFA), only
socio-demographics (admission age, female, emergency admission,
insurance--medicare, race).

Figure \ref{fig:vibration:confounders} shows that small
deviation from the ignorability assumptions is tolerable: for example, removing
antibiotics does not impact the estimate. However, the larger the
deviation from graph \nameref{apd:causal_diagram_albumin}, the larger the bias
compared to the gold-standard. Adjusting only for socio-demographics features
is the closest from an unadjusted risk difference, indicating that we lack
important confounders on the patient health state. This stability of the
treatment effect estimator once sufficient confounders have been included has
already been described and suggested as a confounder selection method
\cite{loh2021confounder}.
\clearpage

\paragraph*{S9 Fig.}
\label{apd:vibration_analysis_for_aggregation}
{\bf Vibration analysis for aggregation.}

We conducted a dedicated vibration analysis on the different choices of
features aggregation, studying the impact on the estimated ATE. We also
studied if some choices of aggregation led to substantially poorer overlap.

We assessed overlap with two different methods. As recommended by
\cite{austin2015moving}, we did a graphical assessment by plotting the
distribution of the estimated. The treatment model hyper-parameters were chosen
by random search, then predicted propensity scores were obtained by refitting
this estimator with cross-fitting on the full dataset.

\begin{figure}[h!]
  \centering
  \begin{subfigure}[b]{0.45\linewidth}
    \centering
    \includegraphics[width=\linewidth]{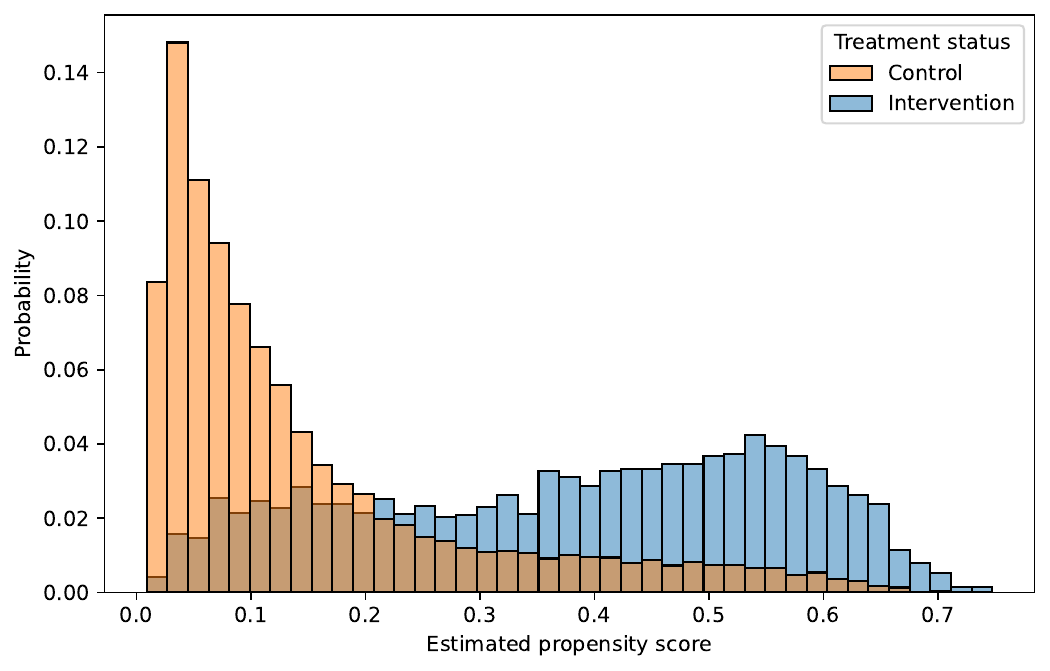}
    \caption{}\label{apd:fig:albumin_for_sepsis:overlap_measure_first_last_median}
  \end{subfigure}
  \hfill
  \begin{subfigure}[b]{0.45\linewidth}
    \centering
    \includegraphics[width=\linewidth]{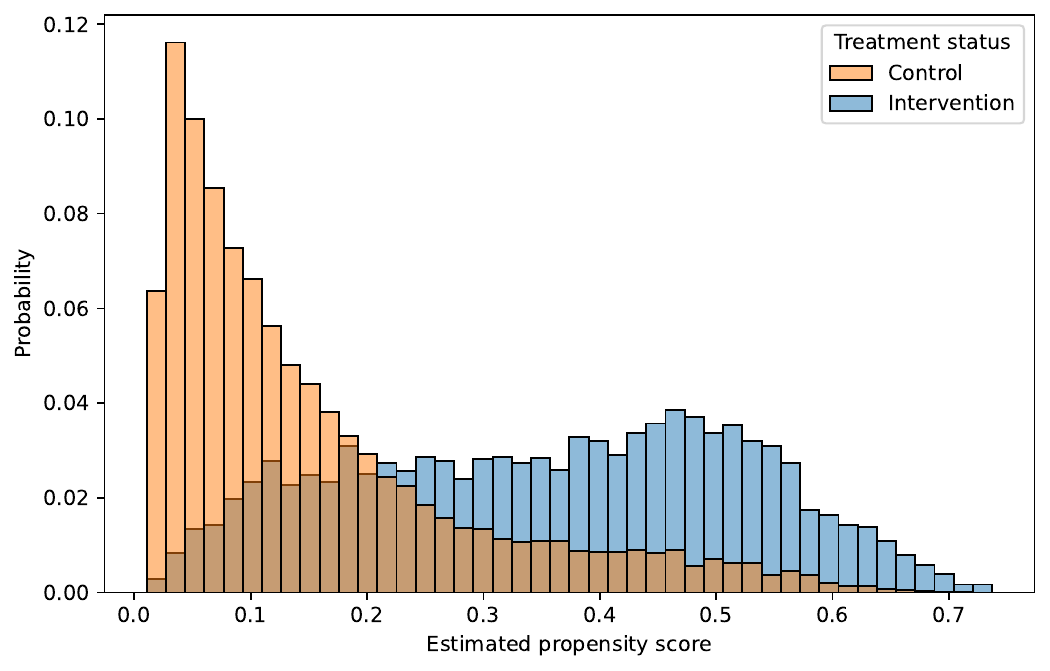}
    \caption{}\label{apd:fig:albumin_for_sepsis:overlap_measure_first}
  \end{subfigure}
  \caption{Different choices of aggregation yield qualitatively close
    distributions of the propensity score: Figure
    \ref{apd:fig:albumin_for_sepsis:overlap_measure_first_last_median})a)
    shows a concatenation of first, last and median measures whereas Figure
    \ref{apd:fig:albumin_for_sepsis:overlap_measure_first})b) shows an
    aggregation by taking the first measure only. The underlying treatment
    effect estimator is a random forest. }\label{apd:fig:albumin_for_sepsis:overlap_measure}
\end{figure}

As shown in Figure \ref{apd:fig:albumin_for_sepsis:overlap_measure}, we did not find substantial differences between methods
when plotting graphically the distribution of the estimated propensity score.

\begin{figure}[h!]
  \centering
  \includegraphics[width=1.0\linewidth]{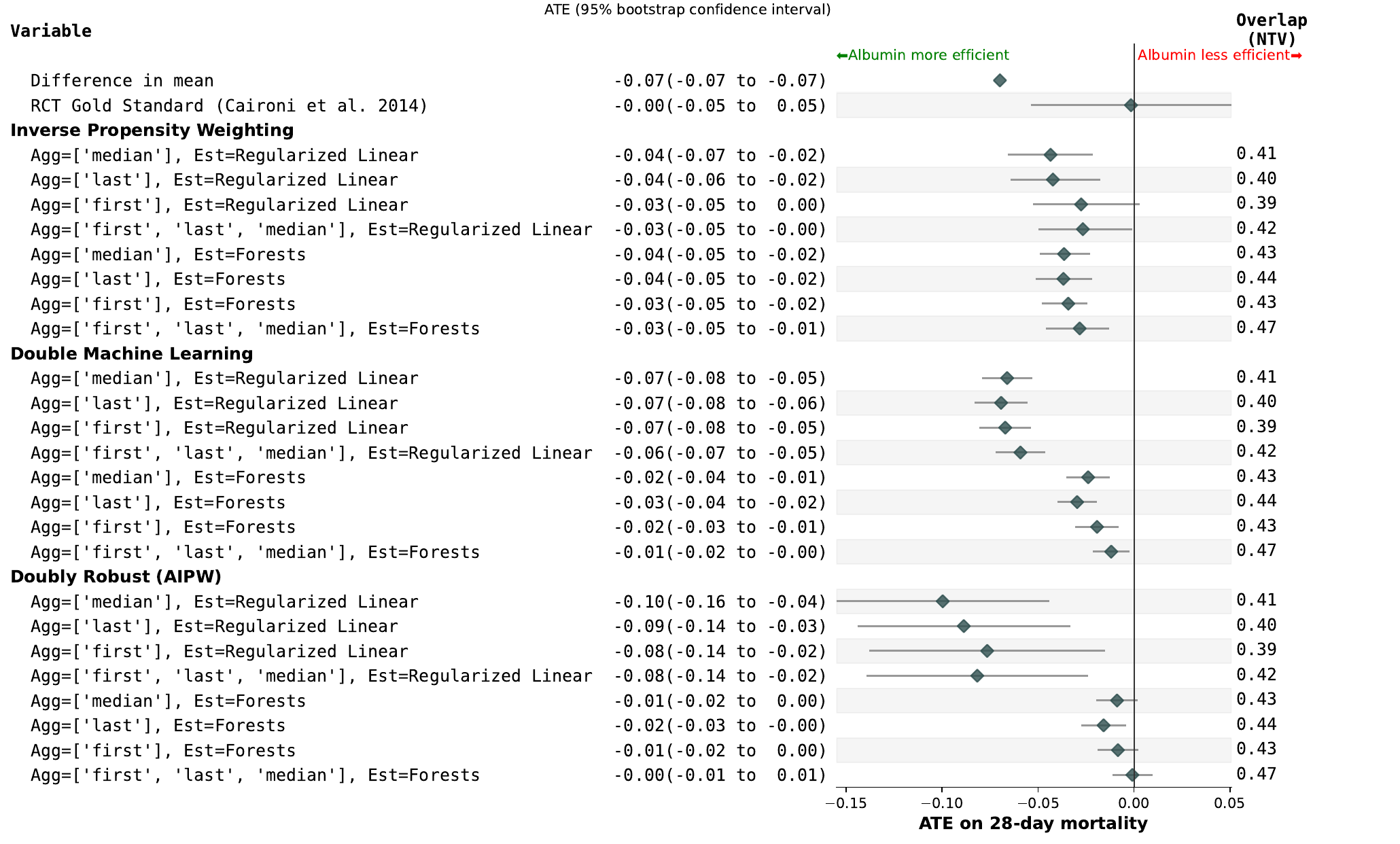}
  \caption{Vibration analysis dedicated to the aggregation choices. The
    choices of aggregation only marginally modify the results. When assessed
    with Normalized Total Variation, the overlap assumption is respected for all
    our choices of aggregation. The green diamonds depict the mean effect and
    the bar are the 95\% confidence intervals obtained by 50 bootstrap
    repetitions.}\label{apd:fig:albumin_for_sepsis:vibration_analysis_for_aggregation}
\end{figure}

We also used normalized total variation (NTV) as a summary statistic of the
estimated propensity score to measure the distance between treated and control
population \cite{doutreligne2023select}. This statistic varies between 0 --
perfect overlap -- and 1 -- no overlap at all. Fig
\ref{apd:fig:albumin_for_sepsis:vibration_analysis_for_aggregation} shows no
marked differences in overlap as measured by NTV between aggregation choices,
comforting us in our expert-driven choice of the aggregation: a concatenation
of first and last feature observed before inclusion time.

\clearpage

\paragraph*{S7 Appendix.}
\label{apd:hte}
{\bf Details on treatment heterogeneity analysis.}
\textbf{Detailed estimation procedure}

The estimation of heterogeneous effect based on Double Machine Learning adds
another step after the computation, regressing the residuals of the outcome
nuisance $\tilde{Y} - \mu(X)$ against the residuals of the treatment nuisance
$\tilde{A} = A - e(X)$ with the heterogeneity features $X_{CATE}$. Noting the
final CATE model $\theta$, Double ML solves:

$$\argmin_{\theta} \mathbb E_n \big[(\tilde{Y} - \tau (X_CATE) \cdot \tilde{A})^2\big ]$$

Where $\tilde{Y} = Y - \hat m(X)$ and $\tilde{A} = A - \hat e(X)$

To avoid the over-fitting of this last regression model, we split the dataset of
the main analysis into a train set (size=0.8) where the causal estimator and the final
model are learned, and a test set (size=0.2) on which we report the predicted Conditional Average
Treatment Effects.

\textbf{Known heterogeneity of treatment for the emulated trial}\label{apd:cate_literature}

\cite{caironi2014albumin} observed statistical differences in the post-hoc
subgroup analysis between patient with and without septic shock at inclusion.
They found increasing treatment effect measured as relative risk for patients
with septic shock (RR=0.87; 95\% CI, 0.77 to 0.99 vs 1.13;95\% CI, 0.92 to 1.39).

\cite{safe2007saline} conducted a post-hoc subgroup analysis of patients with or
without brain injury --defined as Glasgow Coma Scale between 3 to 8--. The
initial population was patients with traumatic brain injury (defined as history
or evidence on A CT scan of head trauma, and a GCS score <= 13). They found
higher mortality rate at 24 months in the albumin group for patients with severe
head injuries.

\cite{zhou2021early} conducted a subgroup analysis on age (<60 vs >60), septic
shock and sex. They conclude for increasing treatment effect measured as
Restricted Mean Survival Time for Sepsis vs septic shock (3.47 vs. 2.58), for
age >=60 (3.75 vs 2.44), for Male (3.4 vs 2.69). None of these differences were
statistically significant.

\textbf{Vibration analysis}\label{apd:cate_results}

\begin{figure}[h!]
  \centering
  \includegraphics[width=0.8\linewidth]{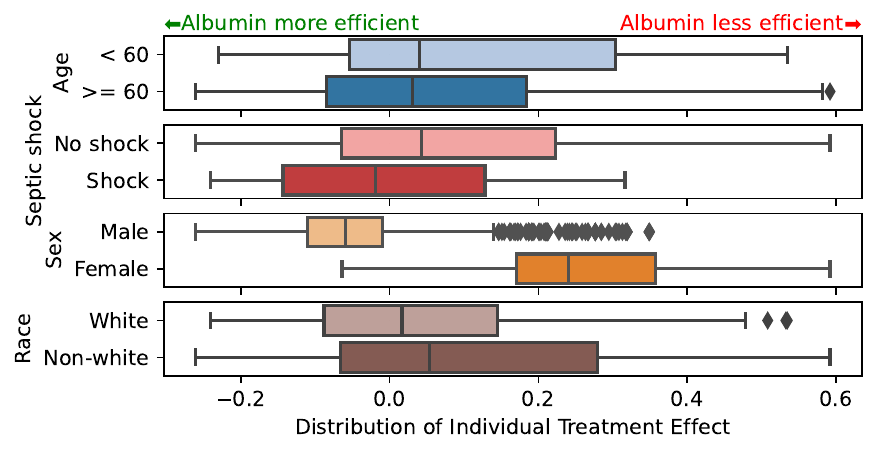}
  \caption{Distribution of Conditional Average Treatment effects on sex, age,
    race and pre-treatment septic shock estimated with a final forest
    estimator. The CATE are positive for each subgroups, which is not
    consistent with the null treatment effect obtained in the main analysis.
    The boxes contain between the 25th and 75th percentiles of the CATE
    distributions with the median indicated by a vertical line. The whiskers
    extends to 1.5 the inter-quartile range of the
    distribution.}\label{apd:fig:albumin_for_sepsis:cate_boxplot_forest}
\end{figure}

\begin{figure}
  \centering
  \includegraphics[width=\linewidth]{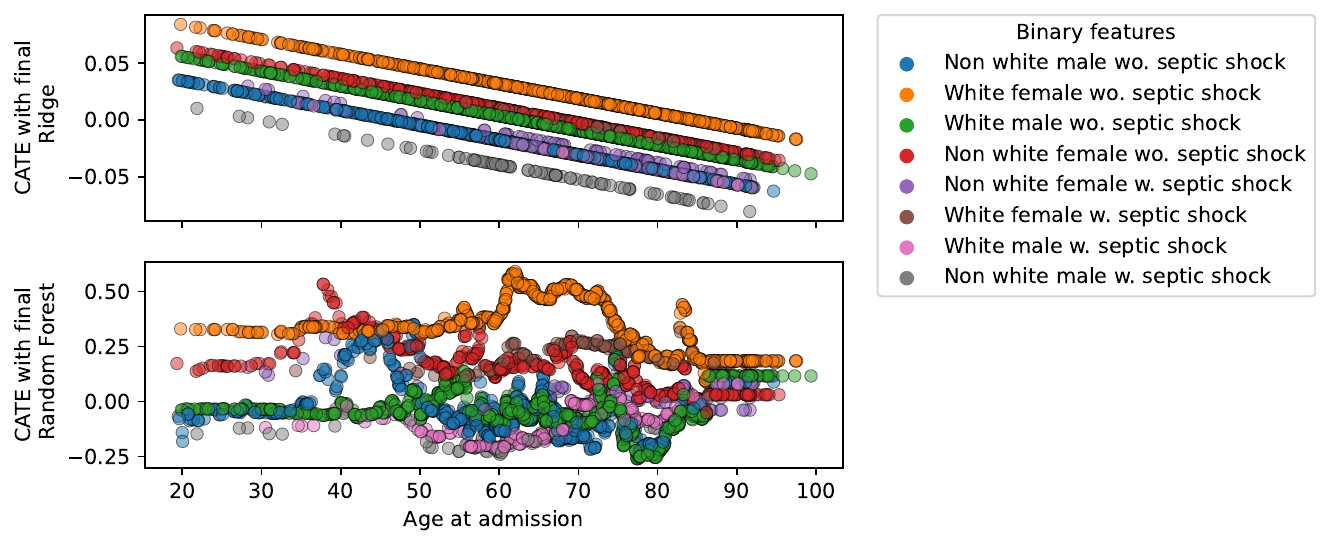}
  \caption{Distribution of Conditional Average Treatment effects on sex, age,
    race and pre-treatment septic shock plotted for different ages. On the top
    the final estimator is a linear model; on the bottom, it is a random
    forest. The forest-based CATE displays more noisy trends than the
    linear-based CATE. This suggest that the flexibility of the random forest
    might be underfitting the data.}\label{apd:fig:albumin_for_sepsis:cate_age_forest_failure}
\end{figure}

The choice of the final model for the CATE estimation should also be informed
by statistical and clinical rationals. Figure
\ref{apd:fig:albumin_for_sepsis:cate_boxplot_forest} shows the distribution of
the individual effects of a final random forest estimator, yielding CATE
estimates that are not consistent with the main ATE analysis. Figure
\ref{apd:fig:albumin_for_sepsis:cate_age_forest_failure} shows that the choice
of this final model imposes an inductive bias on the form of the heterogeneity
and different sources of noise depending of the nature of the model. A random
forest is noisier than a linear model. Figure
\ref{apd:fig:albumin_for_sepsis:cate_age_forest_failure} shows the difference
of modelization on the subpopulation of non-white male patients without septic
shock. One can see that the decreasing linear trend is reflected by the
random forest model only for patients aged between 55 and 80.

\begin{figure}
  \centering
  \includegraphics[width=\linewidth]{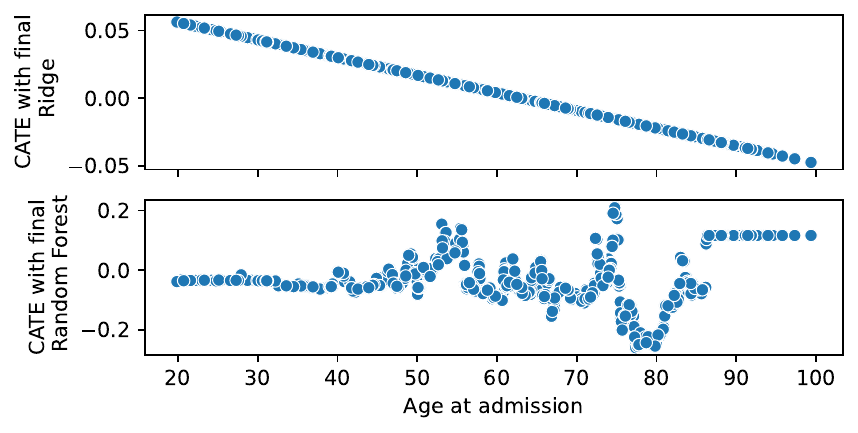}
  \caption{Figure \ref{apd:fig:albumin_for_sepsis:cate_age_forest_failure} on
    the subpopulation of white male patients without septic shock. Contrary to
    the ridge regression (on top) inducing a nicely interpretable trend, using
    random forests as the final estimator failed to recover CATE on ages: the
    predicted estimates do not exhibit any trend and display inconsistently
    large effect sizes, suggesting data underfitting.
  }\label{apd:fig:albumin_for_sepsis:cate_failure}
\end{figure}
\clearpage

\clearpage

%
%
%





\end{document}